\documentclass[useAMS,usenatbib,aas_macros]{mn2e}
\usepackage{graphicx}
\usepackage{amssymb}

\usepackage{epsfig}
\usepackage{color}
\usepackage{amssymb}
\usepackage{ams math}
\usepackage[greek,english]{babel}
\usepackage{subfigure}

\def\nii{[NII]}
\def\ha{H$\alpha$}
\def\hb{H$\beta$} 
\def\oii{[OII]$\lambda$3727}

\def\oiiia{[OIII]$\lambda$4958}
\def\oiiib{[OIII]$\lambda$5007}
\def\oiii{[OIII]$\lambda$$\lambda$4958,5007}

\def\ngal{93 }

\title{The mass-metallicity-star formation rate relation at z $\gtrsim$ 2 with 3D-HST}
\author[F. Cullen, M. Cirasuolo, R.J. McLure, J.S. Dunlop, R.A.A. Bowler]{\parbox\textwidth{F. Cullen$^{1}$\thanks{E-mail:fc@roe.ac.uk}, M. Cirasuolo${^{2,1}}$, R.J. McLure${^1}$, J.S. Dunlop${^1}$, R.A.A. Bowler${^1}$}\\
$^{1}$SUPA\thanks{Scottish Universities Physics Alliance}, Institute for Astronomy, University of Edinburgh, Royal Observatory, Edinburgh EH9 3HJ \\ 
$^{2}$UK Astronomy Technology Centre, Science and Technology Facilities Council, Royal Observatory, Edinburgh, EH9 3HJ}
\begin{document}

\date{Accepted -- . Received 2013 October 3}

\pagerange{\pageref{firstpage}--\pageref{lastpage}} \pubyear{2013}

\maketitle	

\label{firstpage}

%%%%%%%%%%%%%%%%%% ABSTRACT %%%%%%%%%%%%%%%%%%%
\begin{abstract}

We present new accurate measurements of the mass, metallicity and star-formation rate of a statistically-significant-sample of \ngal galaxies at z$\gtrsim$2 using near-infrared spectroscopy taken as part of the 3D-HST survey. We derive a mass-metallicity relation (MZR) for our sample with metallicities based on the oxygen and \hb \ nebular emission lines. We find the MZR derived from our data to have the same trend as previous determinations in the range 0$<$z$<$3 with metallicity decreasing with stellar mass. However, we find that our MZR is offset from a previous determination at z$\gtrsim$2 which used  metallicities derived from the \nii/\ha \ ratio. Incorporating star formation rate information, we find that our galaxies are also offset from the fundamental metallicity relation (FMR) by $\sim$0.3dex. 

Using the Baldwin-Phillips-Terlevich (BPT) diagram we argue that, if the physical conditions of star-forming regions evolve with redshift, metallicity indicators based on [NII] and \ha, calibrated in the local Universe, may not be consistent with the ones based on oxygen lines and \hb. Our results thus suggest that the evolution of the FMR previously reported at z$\sim$2-3 may be an artefact of the differential evolution in metallicity indicators, and caution against using locally-calibrated empirical metallicity relations at high redshift, which do not account for evolution in the physical conditions of star-forming regions.

\end{abstract} 

\begin{keywords}
galaxies: evolution - galaxies: high redshift - galaxies: star-formation -
galaxies: fundamental parameters
\end{keywords}

%%%%%%%%%%%%%%%%% INTRODUCTION %%%%%%%%%%%%%%%%%

\section{Introduction}

Over recent years evidence has been accumulating for the existence of a tight relationship between mass, gas-phase metallicity (hereafter metallicity) and star-formation rate (SFR) in galaxies out to z $\sim$ 2.5. This so called `fundamental metallicity relation' (FMR) was first proposed by \citet{mannucci10} and \citet{lara_lopez10}, though the SFR dependence of the mass-metallicity relation (MZR) had been initially noted earlier by \citet{ellison08}. The existence of an FMR has since been supported by other studies \citep[e.g.][]{cresci11, richard11, mannucci11, lara_lopez12, niino12, belli13, stott13, henry13a, henry13b}, though the exact form of the FMR has been shown to be dependent on the methodology adopted \citep[e.g.][]{yates12, andrews_martini13}. In their original paper, \citet{mannucci10} proposed a FMR which defines a surface in the three dimensional mass-metallicity-SFR space, and in \citet{mannucci11} this original FMR was extended to lower masses. Local Solan Digital Sky Survey (SDSS) galaxies lie on the FMR surface with a residual dispersion of $\sim$ 0.05 dex in metallicity, and their FMR defined locally is seen to extend out to z $\sim$ 2.5 \citep{mannucci10}. This suggests that the physical processes acting on star-forming galaxies have been consistent over the past $\sim$ 10 Gyr of cosmic time. Interestingly, however, \citet{mannucci10} observed an evolution of $\sim$ 0.6 dex away from the FMR for galaxies at z $>$ 2.5, suggesting some change in the processes acting in star-forming galaxies at this epoch.

The original motivation for investigating the relationship between mass, metallicity and SFR was as an attempt to explain some of the features of the more extensively studied MZR. The MZR has been established across a large range in redshift from the local Universe \citep{tremonti04, kewley08, liu08, panter08}, to z $\sim$ 1 \citep{savaglio05, shapley05, cowie08, rodrigures08, zahid11, roseboom12, zahid13}, z $\sim$ 2 \citep{erb06b, henry13b, kulas13}, and out to z $\sim$ 3  \citep{maiolino08, mannucci09}. A consistent picture has emerged from these studies of an MZR similar in shape but differing in normalization out to high redshifts. At a given redshift, galaxies with lower stellar mass are found to have lower metallicities and at a given stellar mass, galaxies at higher redshift are found to have lower metallicities. In the context of the FMR, the evolution of the MZR with redshift is explained due to the fact that  metallicity anti-correlates with SFR, and at higher redshifts galaxies of a given mass have higher SFRs \citep[e.g.][]{daddi07}.

Studying the MZR and FMR requires an accurate measure of galaxy metallicity. One outstanding issue with the observational measurements of metallicities at high redshifts is that the only feasible method is to use strong nebular emission lines. These lines are not direct tracers of metallicity and therefore have to be calibrated in some way. One method that has been adopted is to empirically calibrate these emission line ratios in the local Universe. At low metallicities the line ratios are calibrated using direct metallicity tracers such as the weak [OIII]$\lambda$4363 auroral line, whereas at high metallicities, the auroral line becomes too weak and instead local galaxy spectra are fitted with photo-ionization models of star-forming regions to calibrate the line ratios \citep[see e.g.][]{nagao06,maiolino08}. Using these local empirical calibrations at high redshift assumes that the calibrations sufficiently account for the change in physical conditions of star-forming galaxies with redshift. However, from data currently available there is evidence to suggest that conditions in star-forming regions are evolving with redshift \citep[e.g.][]{brinchmann08, hainline09, shirazi13, kewley13a,kewley13b, nakajima_ouchi13}, and if this is the case using empirical metallicity calibrations at high redshifts may not be reliable. \footnote{It is important to note that purely theoretical metallicity calibrations \citep[e.g.][]{kewley02} are in principal not subject to the potential bias introduced by using local Universe galaxies to calibrate emission line ratios, as theoretical models can explore a large range in ionization conditions which may not be sampled in local galaxies. However, the FMR is built upon the local empirical metallicity calibration of \citet{maiolino08} and so this calibration will be used throughout this paper.}

	\begin{figure}
	\centerline{\includegraphics[width=\columnwidth]{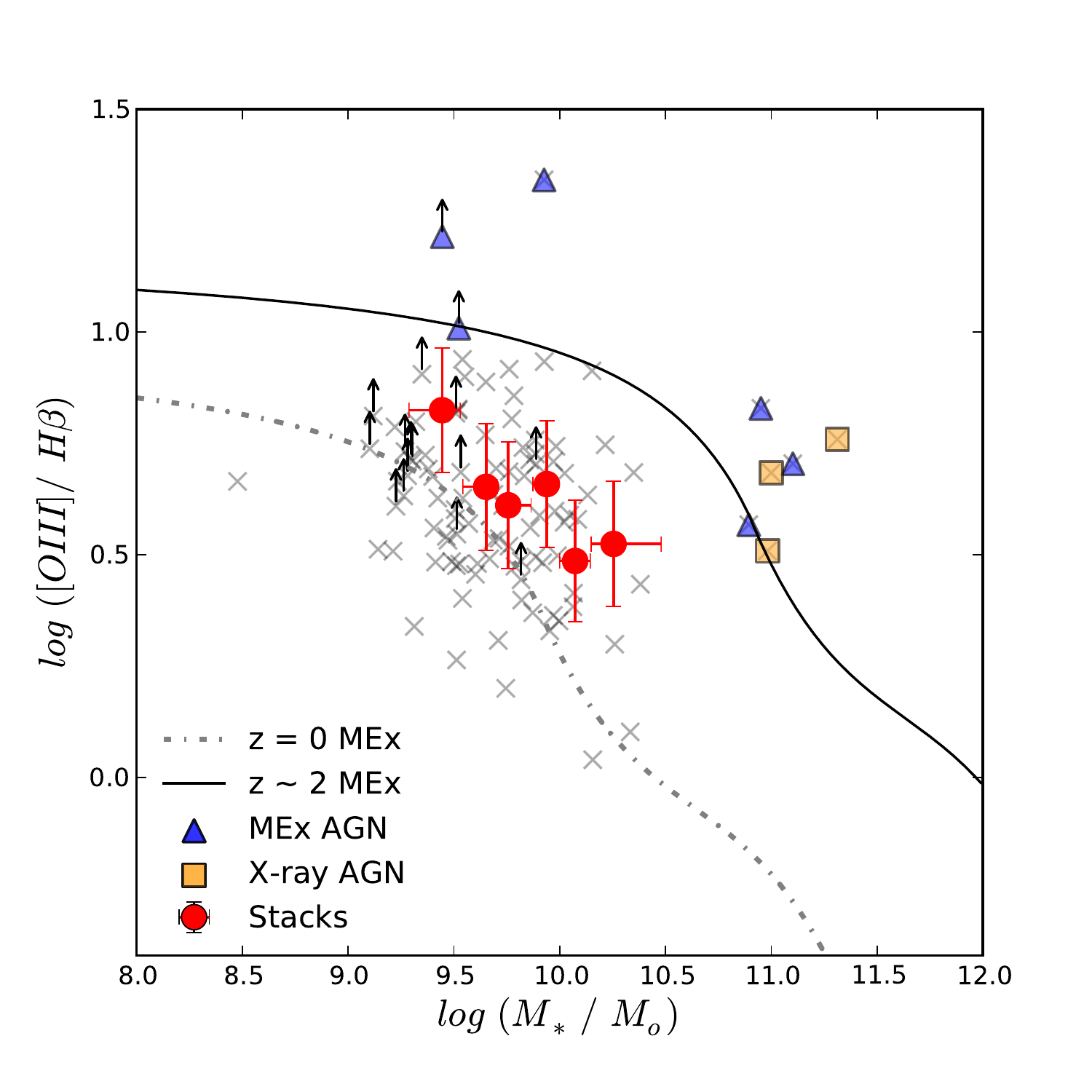}}
	\caption{A modified version of the MEx diagram \citep{juneau11} used for selecting AGN in emission line galaxies. The original form of the separation line given in \citet{juneau11} is given by the grey dashed line and the black solid line shows the modified version of this line used for our z $\gtrsim$ 2 sample (see Sec. \ref{sec_final_sample} for discussion). The grey crosses show individual galaxies in the sample which are not classified as AGN. The yellow squares show AGN identified from x-ray counterparts and the blue triangles show MEx classified AGN. The black arrows represent individual galaxies with only an upper limit on the \oiiib/\hb \ ratio. The red circles with error bars show the position in the MEx diagram of the final stacked galaxies.}
	\label{fig_mex}
	\end{figure}

	\begin{figure}
	\centerline{\includegraphics[width=\columnwidth]{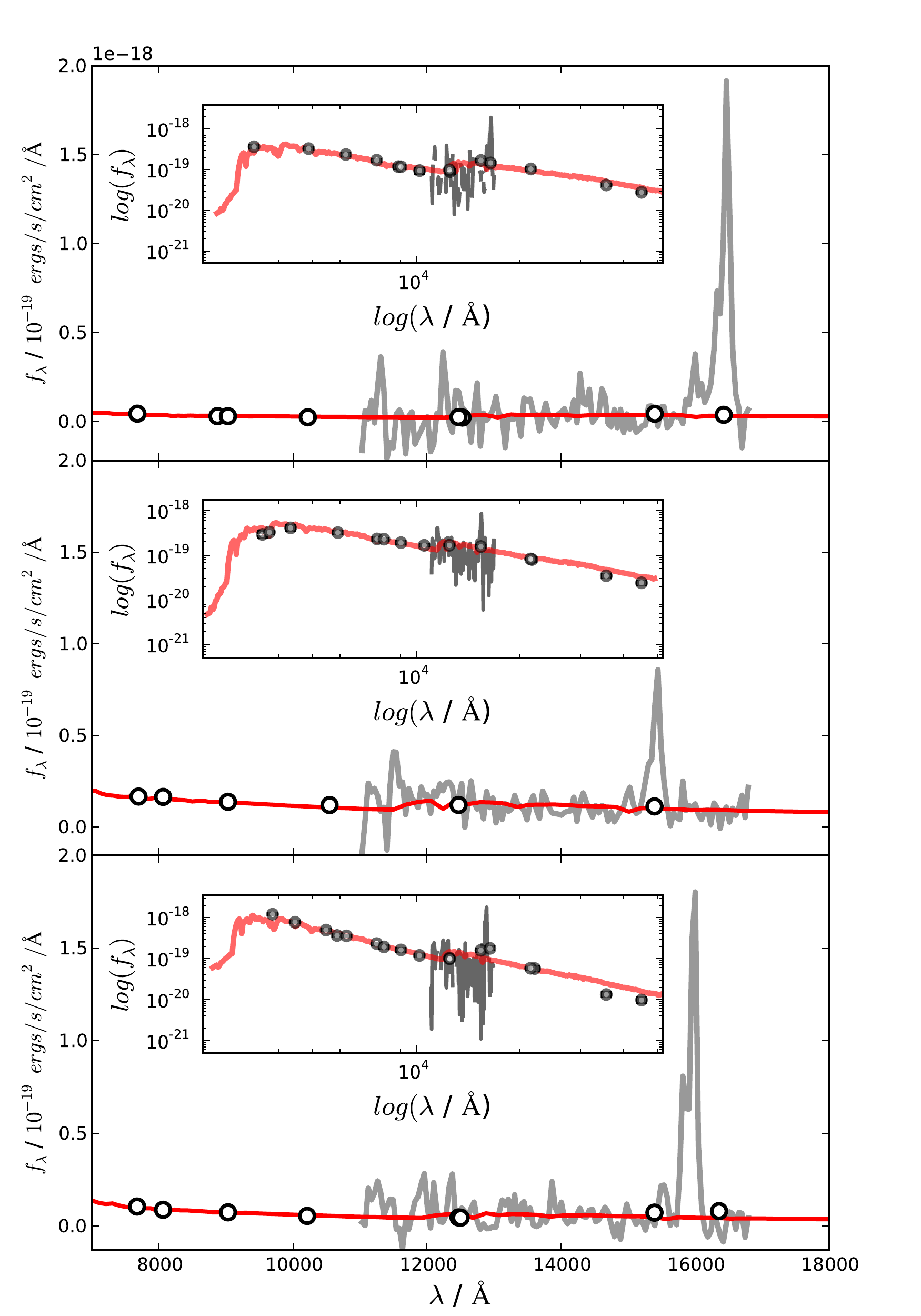}}
	\caption{The continuum fits to three galaxies in our sample. Each panel shows a grism spectra in the rest frame with the continuum fit from the SED plotted in red and the photometry points plotted as black empty circles. Inset in each panel shows the full best-fitting SED of the galaxy with the photometry and grism spectra over plotted.}
	\label{fig_continuum_fit}
	\end{figure}

The evolution of the physical conditions in star-forming regions with redshift is inferred from studying the relationships between nebular emission lines. One common diagnostic is a plot of the \nii/\ha \ versus \oiiib/\hb \ line ratios, known as a BPT diagram \citep*{baldwin81}. Star-forming galaxies in the local Universe form a tight sequence in the BPT diagram \citep[e.g.][]{kauffmann03}, however an evolution away from this sequence is observed at high redshifts towards higher \oiiib/\hb \ and \nii/\ha \ ratios \citep[e.g][]{shapley05, erb06b, hainline09, yabe12}. At the same time an evolution in the ionization parameter of galaxies, which is a measure of the degree of excitation of HII regions, is observed, in that higher redshift star-forming galaxies are found to have higher ionization parameters \citep[e.g.][]{lilly03, hainline09, richard11, nakajima13}. This evolution towards higher ionization parameters has been used to explain the offset of high redshift star-forming galaxies from the local Universe relation in the BPT diagram \citep[e.g.][]{brinchmann08,kewley13a,kewley13b}, and has most recently been suggested as a possible reason for the apparent evolution of galaxies away from the FMR at z $\sim$ 3 \citep{nakajima_ouchi13}. Interestingly, using a sample of low-redshift galaxies with elevated ionization parameters, supposed analogues of these high redshift star-forming galaxies, \citet{ly14} observe an offset from the FMR of the order of 0.1 - 1 dex similar to that observed at z $\sim$ 3.

In this paper, we aim to provide additional observational constraints on the MZR and FMR at $z$ $\gtrsim$ 2, using a sample of star-forming galaxies with a mean redshift $\langle z \rangle$ = 2.16 selected from the 3D-$HST$ near-infrared (near-IR) spectroscopic data set. 3D-HST \citep{brammer12_3dhst} is a slitless spectroscopic survey with the Wide-Field Camera 3 (WFC3) on the Hubble Space Telescope ($HST$) and provides an ideal data set for obtaining a large sample of $z$ $\gtrsim$ 2 galaxies for which metallicities can be measured as the \oii, \hb \ and \oiii \ nebular emission fall into the grism spectra in the redshift range 2 $<$ z $<$ 2.3. We provide a sample of galaxies similar in size to the current largest dataset of z $\gtrsim$ 2 galaxies with measured metallicities studied by \citet{erb06b} ($\langle z \rangle$ = 2.26). Since the \citet{erb06b} study was based on a sample of ultraviolet (UV) selected star-forming galaxies observed from the ground, our line-flux limited sample taken with the HST allows us to probe lower SFRs than the \citet{erb06b} data. Finally, since \citet{erb06b} measure metallicities with the \ha \ and \nii \ nebular lines, we are able to use an independent method with the \oii, \hb \ and \oiii \ lines. These are the same nebular emission lines as used for measuring metallicities in the z $\sim$ 3 studies \citep{maiolino08, mannucci09}, and therefore our data provide useful insight into the observed shift of the FMR between z $\sim$ 2 and z $\sim$ 3.

This paper is organized as follows: in Section \ref{data} we describe the spectroscopic and photometric data, and the measurement of stellar mass, SFRs and metallicities from these data. In Section \ref{results} we describe the results of this paper in the context of the MZR, FMR and photoionization conditions of the galaxies in our sample. We compare our results to those from previous z $\sim$ 2 studies and summarize our results in Section \ref{discussion}. Throughout, we assume a cosmology with $\Omega_m$ = 0.3, $\Omega_{\Lambda}$ = 0.7 and $H_0$ = 70 km s$^{-1}$ Mpc$^{-1}$.

%%%%%%%%%%%%%%%%%%%%%%%%%%% DATA  %%%%%%%%%%%%%%%%%%%%%%%%%%%%%%%%%%%%%%%%%

\section{The Data}\label{data}

In this paper we take advantage of two complimentary data sets: a spectroscopic sample drawn from the 3D-$HST$ spectroscopic survey \citep{brammer12_3dhst} and photometric data taken as part of the CANDELS survey \citep{grogin11,koekemoer11}. The combination of these data allows us to calculate the required physical properties for the galaxies in our sample.

%%%%%%%%%%%%%%%%%%%%%%%% SPECTROSCOPY %%%%%%%%%%%%%%%%%%%%%%%%%%%%%%%%%%%%

\subsection{3D-$HST$ Spectroscopic Data}

	\begin{figure*}
	\centerline{\includegraphics[width=7in]{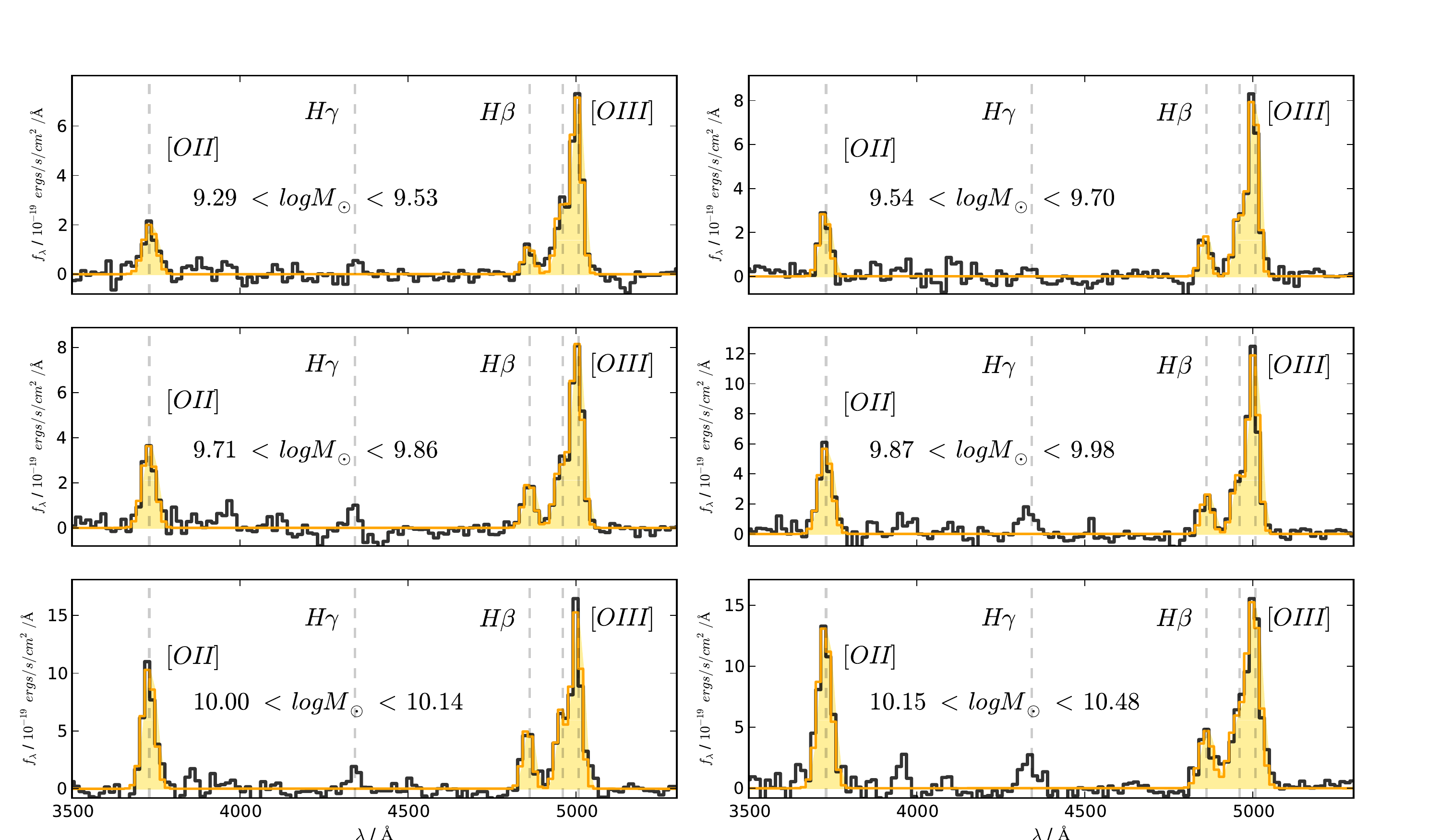}}
	\caption{The rest-frame stacked spectra of galaxies in our sample split into six bins of increasing stellar mass. Each panel is labeled with the range of stellar mass in each bin and the \oii , \hb \ and blended \oiii \ emission lines are marked by dashed lines. The stacked spectra are shown in black with the fit to the emission lines overlaid in orange.}
	\label{fig:spectra}
	\end{figure*}

The spectroscopic data used in this paper are part of the 3D-$HST$ observations described in \citet{brammer12_3dhst}. The 3D-$HST$ survey provides low resolution (R $\sim$ 130) spatially resolved near-IR grism spectra over the wavelength range 1.1 - 1.68$\mu$m taken with the WFC3 G141 grism on the $HST$. The survey covers 675 arcmin$^2$ providing spectroscopic follow-up of three quarters of the deep near-IR CANDELS imaging in five fields (AEGIS, COSMOS, GOODS-S, GOODS-N and UDS).

A modified version of the publicly available pipeline described in \citet{brammer12_3dhst} was used to reduce the 3D-$HST$ grism exposures utilizing the \textsc{aXe} software package \citep{kummel09}. Objects are selected from a CANDELS $F$160W ($H$$_{160}$) mosaic of the given field down to a limiting magnitude of $H$$_{160}$ = 26 (AB mag). The four raw direct images of the 3D-$HST$ pointing are combined using the \textsc{iraf} task \textsc{multidrizzle} and positional offset from the CANDELS mosaic are calculated using \textsc{tweakshifts}. The four raw grism images are combined using \textsc{multidrizzle} and aligned to the reference mosaic using the shifts calculated on the direct image. The standard set of \textsc{aXe} tasks are then used to fully reduce the grism pointings to individual 1D and 2D spectra. A custom method for estimating background subtraction as described in \citet{brammer12_3dhst} is implemented, this takes into account the variation of IR background across the sky which cannot be fully accounted for in default grism master sky background \citep{kummel09}.

We carefully treat the contamination of the grism spectra. With grism spectroscopy any given spectrum can be contaminated by the spectra of nearby sources. To mitigate this effect, in some grism surveys, exposures are taken at different position angles in the hope that multiple dispersion directions reduce the chances of a spectrum being severely contaminated \citep[e.g][]{pirzkal12}. However, grism exposures in the 3D-$HST$ survey are taken at only one position angle, and so a reliable quantitative estimation of spectral contamination is required. We estimate the contamination by using the images in the $F$850LP, $F$125W and $F$160W bands available in the CANDELS fields. We use the \textsc{aXe} task \textsc{fcubeprep} to construct model grism images using the photometric and morphological properties derived from the images; the mosaics are extended around the field of view (FOV) of the grism exposure to account for sources outside the FOV whose spectra are dispersed on to the grism image. In this way a quantitative estimate of the contamination for each spectrum is produced and we subtract the contamination model from the observed spectrum to obtain the final spectrum of each object.

\subsubsection{Redshift determination}

Due to the low resolution of the grism spectra, care must be taken when deriving galaxy redshifts. In many cases only one emission line is visible in the spectrum and so combining spectral redshift information with photometric redshift estimates is required. We estimate redshifts using a combination of spectral template fitting and photometric redshifts. For spectral fitting we use galaxy templates from the K20 survey \citep{mignoli05_k20}, binned to the resolution of the grism spectra and calculate probability distributions derived from a $\chi^2$ fit. The redshift probability distribution from spectra fitting is then combined with a photometric redshift probability distributions calculated from the broad-band photometry of the galaxies using the EAZY code \citep{brammer08_eazy}.

\subsubsection{Final spectroscopic sample}\label{sec_final_sample}

The sample used in this paper is drawn from three of the five fields covered by 3D-$HST$: GOODS-S, COSMOS and UDS. Within a wavelength range of the grism the prominent \oii, \hb \ and \oiii \ emission lines are observable at 2.0 $<$ z $<$ 2.3. These emission lines can be used for estimating SFRs and metallicities of galaxies as described in \citet{maiolino08} and also allow accurate redshift determination.  The initial sample is assembled from a visual inspection of 3D-$HST$ spectra to identify emission line galaxies in the correct redshift range. We do not exclude contaminated spectra from our sample unless the contamination is severe or the contamination model clearly erroneous (e.g leaving residual continuum features after subtraction). This initial sample contains 103 galaxies and extends down to a flux limit in \oiiib \ of 5 x 10$^{-18}$ $ergs/cm^{2}/s$.

We remove possible active galatic nuclei (AGN) for our sample by first cross-matching with the $Chandra$ 4MS X-ray catalogue in GOODS-S \citep{xue11} and $Chandra$ 1.8Ms X-ray catalogue in COSMOS \citep{civano12}; three galaxies were identified and removed based on the X-ray data. We also followed the method of \citet{henry13b} in using the Mass-Extinction (MEx) diagram \citep{juneau11} to separate star-forming galaxies and AGN based on their \oiiib/\hb \ ratio and mass (Fig. \ref{fig_mex}). We follow \citet{henry13b} in shifting the \citep{juneau11} relation by 1 dex in mass to account for the mass-metallicity evolution at high redshift, we also shift by + 0.2 dex in \oiiib/\hb \ to account for the evolution of the \oiiib/\hb \ ratio in star-forming galaxies at $z$ $\sim$ 2 due to evolving ionization conditions (see Section \ref{sec_photoion} and \citet{kewley13a}). Across the range of \nii/\ha \ occupied by local star-forming galaxies, +0.2 dex is the average shift in the \oiiib/\hb \ ratio between $z$ $\sim$ 0 and $z$ $\sim$ 2 based on the \citet{kewley13b} models. We note that \hb \ is not detected in all individual galaxies and in this case we can only put a lower limit on the \oiiib/\hb \ ratio, however we do not exclude galaxies with lower limits which fall below the AGN separation line. Using this modified MEx diagram approach we remove a further six galaxies from the sample, leaving a total of 9/103 galaxies (9$\%$) identified as AGN. This AGN fraction is consistent with other studies at slightly lower redshift (z $\sim$ 1.5) \citep[e.g.][]{stott13,zahid13}.

Finally one further galaxy is excluded from the stacking analysis (see Section \ref{sec_stack}) as the measured mass of log($M_*/M_o$) = 8.63 (see Section \ref{sec_mass}) is offset by 0.66 dex from the next lowest mass galaxy in the sample and would therefore skew the mass distribution in the lowest mass bin. This leaves a final sample of 93 galaxies on which the following analysis is performed.

\subsection{CANDELS Photometric Data}

The photometric data are taken from three separate catalogues for the GOODS-S, UDS and COSMOS fields. All photometry covers the rest-frame UV to mid-IR ($Spitzer$/IRAC 3.6 and 4.5$\mu m$). In both GOODS-S and UDS we take the photometry from existing CANDELS catalogues described in \citet{guo13} and \citet{galametz13} respectively. For COSMOS we produce a new catalogue by first convolving all the available ground-based and space-based optical and near-IR imaging (see \citet{bowler12} for details) to a common PSF (0.8$^{\prime\prime}$) and then performing aperture photometry using the smoothed F160W image as the detection band. The $Spitzer$/IRAC 3.6 and 4.5$\mu m$ photometry was included by de-confusing the IRAC images using the UltraVISTA $Ks$-band imaging \citep{mccracken12}.

\subsubsection{Stellar masses and SFRs from SED fitting}\label{sec_mass}

Stellar masses and SFRs for individual galaxies are measured using the publicly available code \textsc{lephare} \citep{ilbert06}. We run \textsc{lepahre} with solar-metallicity \citet{bc03} templates assuming $\tau$ model star-formation histories with $\tau$ = 0.3, 1, 2, 3, 5, 10 , 15, 30 Gyr and a Chabrier initial mass function (IMF). The \citet{calzetti00} attenuation law is used to account for dust extinction with $E(B-V)$ values ranging from 0 to 0.6. The age of the model is allowed to vary between 0.05 Gyr and the age of the Universe at the spectroscopic redshift of the galaxy. The normalization factor required to scale the best-fitting template to the observed magnitudes in each band gives the stellar mass. In order to compare to \citet{erb06b} data we correct our masses to the total mass of stars formed using the conversions in the BC03 templates, the median increase in mass when applying this correction is + 0.18 dex.

\subsection{Galaxy Metallicities and SFRs from Spectra}

\subsubsection{Galaxy stacking procedure}\label{sec_stack}

We measure spectral properties (SFR and metallicity) for all the individual galaxies. However, due to the low signal-to-noise ratio (S/N) of some of the spectra we also stacked the spectra in bins of stellar mass, following the method of \citet{erb06b}. The mass bins were chosen such that each bin contained a similar number of galaxies. To produce the stacks each individual spectrum and best-fitting BC03 model were first shifted to the rest frame and de-reddened using the \citet{calzetti00} attenuation law. The continuum of each galaxy was then subtracted by interpolating the best-fitting BC03 model SED on to the same wavelength grid and normalizing to the galaxy spectra with the emission lines masked,  in this way the continuum could be subtracted whilst accounting for the reddening in the emission lines assuming E(B-V)$_{stars}$ $\sim$ E(B-V)$_{gas}$ (see Section \ref{sfr_hb}). Examples of continuum fits to galaxies are shown in Fig. \ref{fig_continuum_fit}. This method of continuum subtraction was preferred to fitting a simple low order polynomial to the galaxy spectra as the SED model can better fit the 4000$\textup{\AA}$ break crossing the \oii \ emission line and also accounts for the underlying \hb \ absorption in the stellar continuum. Within each mass bin the galaxy spectra were interpolated to a common wavelength grid and we determine the median flux within each wavelength bin. The final stacked galaxy spectra are shown in Fig. \ref{fig:spectra}. For consistency we have checked that removing the continuum from each galaxy individually before stacking, and removing a stack of SED models from a stack of the observed galaxy spectra does not change our results. 

We note that stacking the intrinsic fluxes of each galaxy may be biasing our measurements towards the brightest emission lines galaxies in the stack \citep{dominguez13, henry13b}, although taking the median flux in each wavelength bin should mitigate this effect. We have checked that normalizing each individual spectrum by its \oiiib \ flux before stacking does not change our results. We find that when the spectra are normalized the measured metallicities are fully consistent with those measured without normalizing. We use the intrinsic stacked spectra in the remainder of this paper so that we can measure the SFR and metallicity directly from the same spectra.

\subsubsection{Line fluxes}\label{sect_line_flux}

We measure line fluxes from the continuum-subtracted grism spectra in the following way. For the \oiii \ doublet and \hb \ lines we fit a triple Gaussian allowing the height of each Gaussian to vary. Since the \oiiib \ line always has the highest S/N we fit the centroid of the \oiiib \ line and fix the centroids of \oiiia \ and \hb \ given the best-fitting \oiiib \ centroid and redshift. We allow the fit to vary the width of the \oiiib \ line; the width of the \oiiia \ and \hb \ lines are fixed to the best-fitting width of the \oiiib \ line. The \oiii \ lines are blended due to the low resolution of the grism spectra and to measure the flux in these lines we used two methods. First we fit two Gaussians and use the best-fitting parameters for each line directly from the fit, secondly we sum the two Gaussians from the triple Gaussian and take the \oiiib \ flux to be 3/4 of the total \citep{storey00}, this follows previous methods for measuring the \oiii \ line fluxes from grism spectra \citep{trump13}. We have checked that both methods are fully consistent and do not affect the results of this paper. For the \oii \ line we fitted a single Gaussian allowing the centroid, line-width and height to vary. Examples of the line fits to the stacked spectra are shown in Fig. \ref{fig:spectra}. Dust corrected line fluxes are given in Table 1. 

\begin{table}
\begin{minipage}{70mm}
\caption{Measured data for the galaxy stacks in our sample.}
	\begin{tabular}{@{}cccccc}
		\hline
		ID & 
		N$_{gal}$$^a$ & 
		$\langle z \rangle$$^b$ & 
		\emph{f}$_{\rm [OIII]5008}$$^c$ & 
		\emph{f}$_{\rm H\beta}$$^c$  & 
		\emph{f}$_{\rm [OII]}$$^c$ \\
		\hline
		1 & 15 & 2.22 & 2.9 $\pm$ 0.3 & 0.5 $\pm$ 0.1 & 1.0 $\pm$ 0.1 \\
		2 & 15 & 2.11 & 3.1 $\pm$ 0.3 & 0.7 $\pm$ 0.1 & 1.2 $\pm$ 0.2 \\
		3 & 16 & 2.15 & 3.5 $\pm$ 0.4 & 0.9 $\pm$ 0.1 & 1.8 $\pm$ 0.2 \\
		4 & 16 & 2.14 & 4.6 $\pm$ 0.5 & 1.1 $\pm$ 0.1 & 2.9 $\pm$ 0.3 \\
		5 & 16 & 2.17 & 6.1 $\pm$ 0.6 & 2.0 $\pm$ 0.3 & 4.8 $\pm$ 0.5 \\
		6 & 15 & 2.20 & 7.7 $\pm$ 0.8 & 2.3 $\pm$ 0.3 & 6.7 $\pm$ 0.7 \\
		\hline

	\end{tabular}
\end{minipage}
\begin{minipage}{70mm}
	\medskip
		$^a$ Number of individual galaxies in the stack. \\
		$^b$ Average redshift of the stack. \\	 
		$^c$ Dust corrected flux in units of 10$^{-17}$ erg/s/cm$^2$ \\
\end{minipage}

\end{table}

\begin{table*}
\begin{minipage}{140mm}
\centering
\caption{Derived data for the galaxy stacks in our sample.}
	\begin{tabular}{@{}cccccc}
		\hline
		ID &
		log(M/$M_{\odot}$)$^a$ &  
		12 + log(O/H)$^b$ & 
		log(SFR$_{\rm H\beta}$ / M$_{\odot}$ $yr^{-1}$)$^c$ & 
		log(SFR$_{SED}$ / M$_{\odot}$ $yr^{-1}$)$^d$ &
		$E(B-V)$$^e$ \\
		\hline
		1 & 9.44$^{+0.09}_{-0.16}$ & 8.10$^{+0.10}_{-0.11}$ & 0.86 $\pm$ 0.16 & 0.57 $\pm$ 0.15 & 0.04 $\pm$ 0.04 \\
		2 & 9.65$^{+0.05}_{-0.11}$ & 8.13$^{+0.13}_{-0.12}$ & 0.97 $\pm$ 0.17 & 0.78 $\pm$ 0.22 & 0.09 $\pm$ 0.04 \\
		3 & 9.75$^{+0.11}_{-0.05}$ & 8.26$^{+0.11}_{-0.09}$ & 1.09 $\pm$ 0.12 & 0.81 $\pm$ 0.36 & 0.08 $\pm$ 0.07 \\
		4 & 9.94$^{+0.04}_{-0.07}$ & 8.30$^{+0.09}_{-0.09}$ & 1.17 $\pm$ 0.14 & 1.28 $\pm$ 0.31 & 0.19 $\pm$ 0.09 \\
		5 & 10.07$^{+0.07}_{-0.07}$ & 8.41$^{+0.09}_{-0.09}$ & 1.47 $\pm$ 0.16 & 1.37 $\pm$ 0.28 & 0.18 $\pm$ 0.07 \\
		6 & 10.25$^{+0.22}_{-0.11}$ & 8.41$^{+0.09}_{-0.08}$ & 1.56 $\pm$ 0.17 & 1.62 $\pm$ 0.43 & 0.23 $\pm$ 0.13 \\
		\hline
	\end{tabular}
\end{minipage}
\begin{minipage}{140mm}
	\medskip
		\raggedright{$^a$ Median mass of the stack, error bars represent the range of stellar masses in each bin.} \\
		\raggedright{$^b$ Metallicities of the stacks derived from the \citet{maiolino08} calibrations.} \\
		\raggedright{$^c$ The SFR of the stacks measured from the H$\beta$ flux corrected for dust extinction.} \\
		\raggedright{$^d$ Mean and standard deviation of SED SFR of the individual galaxies in the stack.} \\
		\raggedright{$^e$ Mean and standard deviation of $E(B-V)$ of the individual galaxies in the stack.} \\
\end{minipage}
\end{table*}

%%%%%%%%%%%%%%%% METALLICITY %% %%%%%%%%%%%%%%%%%

	\begin{figure}
	\centerline{\includegraphics[width=\columnwidth]{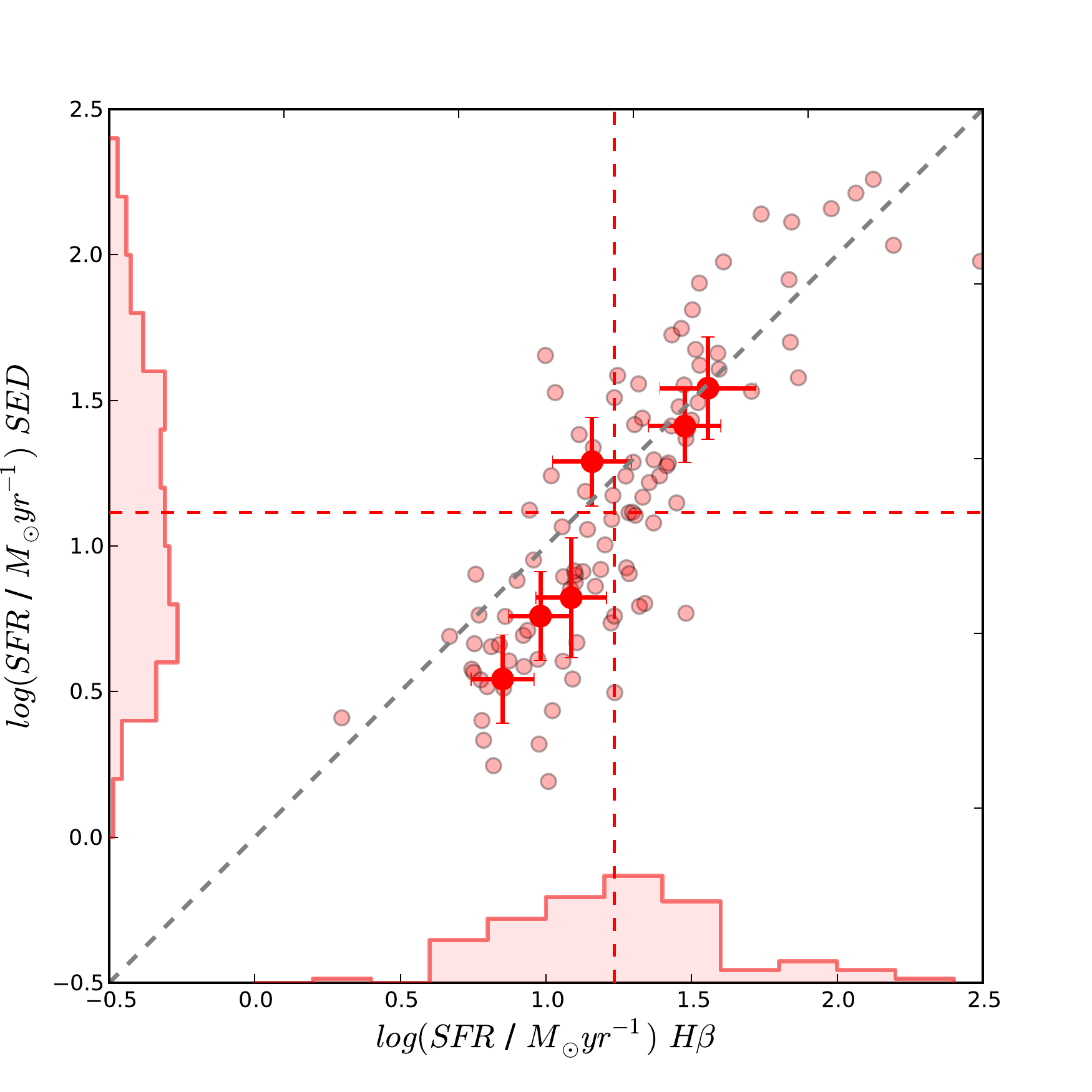}}
	\caption{A comparison of SFR measured from the \hb \ line with SFR as derived from SED fitting. The galaxy stacks are shown as the large red points with error bars. The SED SFRs for the stacks are taken as the median of the SED-derived SFRs of individual galaxies within the stack with measured \hb. The small red points represent the individual galaxies. On both axes histograms of the values for the individual galaxies are plotted and the dotted red lines show the median values for each distribution. Respectively, the median and median absolute deviation (MAD) of the SFRs of the whole galaxy sample are 17.2 $\pm$ 8.8 \textrm{M$_{\odot}$yr$^{-1}$} and 13.0 $\pm$ 12.3 \textrm{M$_{\odot}$yr$^{-1}$}}.
	\label{fig:sfr_comp}
	\end{figure}

%%%%%%%%%%% star-formation %% %%%%%%%%%%%%%%%

\subsubsection{Star-formation rates}\label{sfr_hb}

We measure SFRs from the H$\beta$ line flux, first converting to luminosity and then using the common conversion factor H$\alpha$/H$\beta$ = 2.86 \citep{kennicutt98}. We then scale down by a factor 1.7 to convert to a \citet{chabrier03} IMF. Use of the H$\alpha$/H$\beta$ conversion factor relies on the assumption that the galaxies in the stacks have $E(B-V)=0$ after the de-reddening of individual spectra, therefore a good estimate of the extinction suffered by emission lines in the galaxy is required. There has been much debate in the literature as to whether the extinction returned from SED fitting, which corresponds to the extinction in the stellar continuum, is the same as the extinction suffered by emission lines, since emission lines emanate from dusty star-forming regions and therefore could potentially suffer larger extinction than stars. 

The extra extinction of nebular emission lines is parametrized by a factor $f$ where $E(B-V)_{stars}$=$f*E(B-V)_{gas}$. \citet{calzetti00} find that $f$ = 0.44 $\pm$ 0.03 for local starburst galaxies and similar corrections have been reported for galaxies out to z $\sim$ 2 \citep[e.g.][]{forster06, cresci11, yabe12}. However at intermediate redshifts (z $\sim$ 1.5) there is some evidence for an evolution towards larger $f$ values \citep[e.g.][]{kashino13,price13}, and at z $\sim$ 2 some studies find an apparent evolution to $f$ $\sim$ 1 \citep[e.g.][]{erb06a, hainline09}.

The method of de-reddening spectra described in Section \ref{sec_stack} assumes E(B-V)$_{stars}$ $\sim$ E(B-V)$_{gas}$ so to get a sense as to whether we are making sensible dust corrections we can compare our best-fitting SFR from SED fitting to that derived from the de-reddened line flux. The comparison is shown in Fig. \ref{fig:sfr_comp}. The figure shows that the two independent measures of SFR are in good agreement for the galaxy stacks, for the H$\beta$ and SED methods respectively the median and median absolute deviation (MAD) of the SFRs of the whole galaxy sample are 17.2 $\pm$ 8.8 and 13.0 $\pm$ 12.3 \textrm{M$_{\odot}$yr$^{-1}$}. Given the good agreement between the SFRs, we adopt the $E(B-V)$ values returned by SED fitting for correcting the emission lines therefore following previous studies at z $\sim$ 2 in assuming that E(B-V)$_{stars}$ $\sim$ E(B-V)$_{gas}$.

\subsubsection{Gas phase metallicity}\label{data_metallicity}

The metallicity of a galaxy is a measure of the abundance of metals relative to  hydrogen in the interstellar medium and is most commonly quoted in terms of the oxygen abundance ratio 12 + log(O/H). Measuring the metallicity at high redshifts relies on using ratios of various strong, optical emission lines emanating from HII regions. These line ratios are a combination of strong hydrogen recombination lines such as \ha \ and \hb,  and collisionally excited forbidden transitions in metals such as \nii, \oii \ and \oiii. 

Unfortunately the ratios of the strengths of these lines do not depend solely on the element abundances but have other dependences (e.g. gas density, hardness of ionizing radiation, etc.), and therefore must be calibrated against direct metallicity tracers. At low metallicities, in the local Universe, the electron temperature T$_e$ method \citep{pettini04} can be used to calibrate the nebular lines. At higher metallicities the electron temperature methods cannot be used and we rely on photoionization modelling of HII regions \citep{kewley02}. Various attempts have been made to calibrate these strong nebular line ratios  across a wide range of metallicity values \citep[e.g.][]{nagao06, kewley08, maiolino08}. These authors give various recipes for calculating the metallicity of a galaxy from line ratios of optical emission lines. It is important to note that inferred values of metallicities depend crucially on the calibration adopted, and therefore when making comparisons all results must be converted to a consistent calibration (see \citet{kewley08} for a detailed discussion). In this paper we use the local empirical calibrations described in \citet{maiolino08} as this calibration was used in the original investigation of the FMR \citep{mannucci10}. The \citet{maiolino08} calibrations are derived from a sample for low metallicity galaxies for which metallicities can be measured directly using the T$_{e}$ method, and a sample of SDSS spectra at higher metallicity for which metallicities are derived from photoionization modelling.

\begin{figure}
	\centerline{\includegraphics[width=\columnwidth]{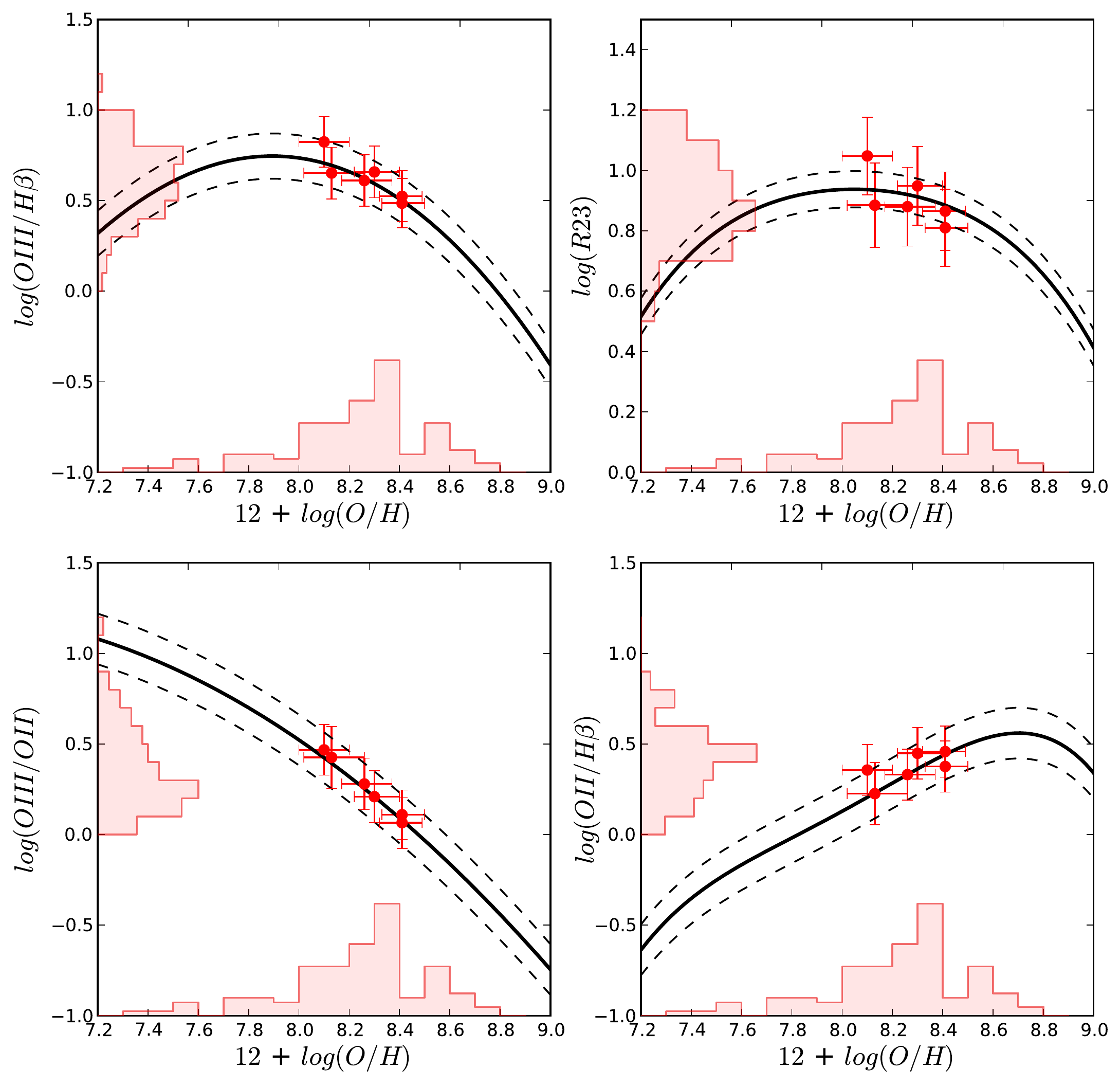}}
	\caption{Relationship between line ratio and derived metallicity values for each of the four metallicity calibrations used to determine galaxy metallicities in this paper. The calibration curves are taken from \citet{maiolino08}. Shown as red solid circles are the metallicities derived using a combination of all four line ratios as described in Section \ref{data_metallicity}. The red histograms represent the distributions of individual galaxies in the sample with detections in the appropriate emission lines.}
	\label{fig_metallicity_curves}
	\end{figure}

We measure metallicities using the \oii, \oiiia, \oiiib \ and \hb \ emissions lines via a method similar to that used for the z $\sim$ 3 galaxies in the AMAZE and LSD surveys \citep{maiolino08, mannucci09}. However, in contrast to these studies, we do not fit to metallicity and extinction simultaneously across all line ratios. However, both \citet{maiolino08} and \citet{mannucci09} note that their fit, whilst constraining the metallicity, does not simultaneously provide good constraints on the dust extinction, implying the derived metallicity is weakly dependent on the value of dust extinction adopted. We measure all line ratios available which have metallicity calibrations given in \citet{maiolino08} (fig. 5 in their paper). Specifically the line ratios are R$_{23}$ (=\oii \ + \oiiia \ + \oiiib /\hb), \oiiib /\hb, \oiiib /\oii \ and \oii /\hb. The line ratios and calibration curves are shown in Fig. \ref{fig_metallicity_curves}. 

To measure the metallicity of a galaxy stack we first calculate the observed line ratio in each of the four diagrams. At each value of metallicity, within a large range (7 $<$ log(O/H) + 12 $<$ 9.5), we calculate, for each calibration curve, a $\chi^2$ statistic from the difference between the observed line ratio and calibration line ratio at that metallicity. Thus for all four calibration curves we construct a $\chi^2$ versus log(O/H) + 12. We combine the $\chi^2$ values from each of the four diagrams to construct an overall $\chi^2$ versus log(O/H) + 12. We take the metallicity at the minimum of the combined $\chi^2$ as the best-fitting metallicity. We take the $1\sigma$ confidence intervals to be within $\Delta\chi^2 = 1$ of the best-fitting value. The metallicities of the stacked spectra along with all other derived properties are given in Table 2.

\begin{figure}
	\centerline{\includegraphics[width=\columnwidth]{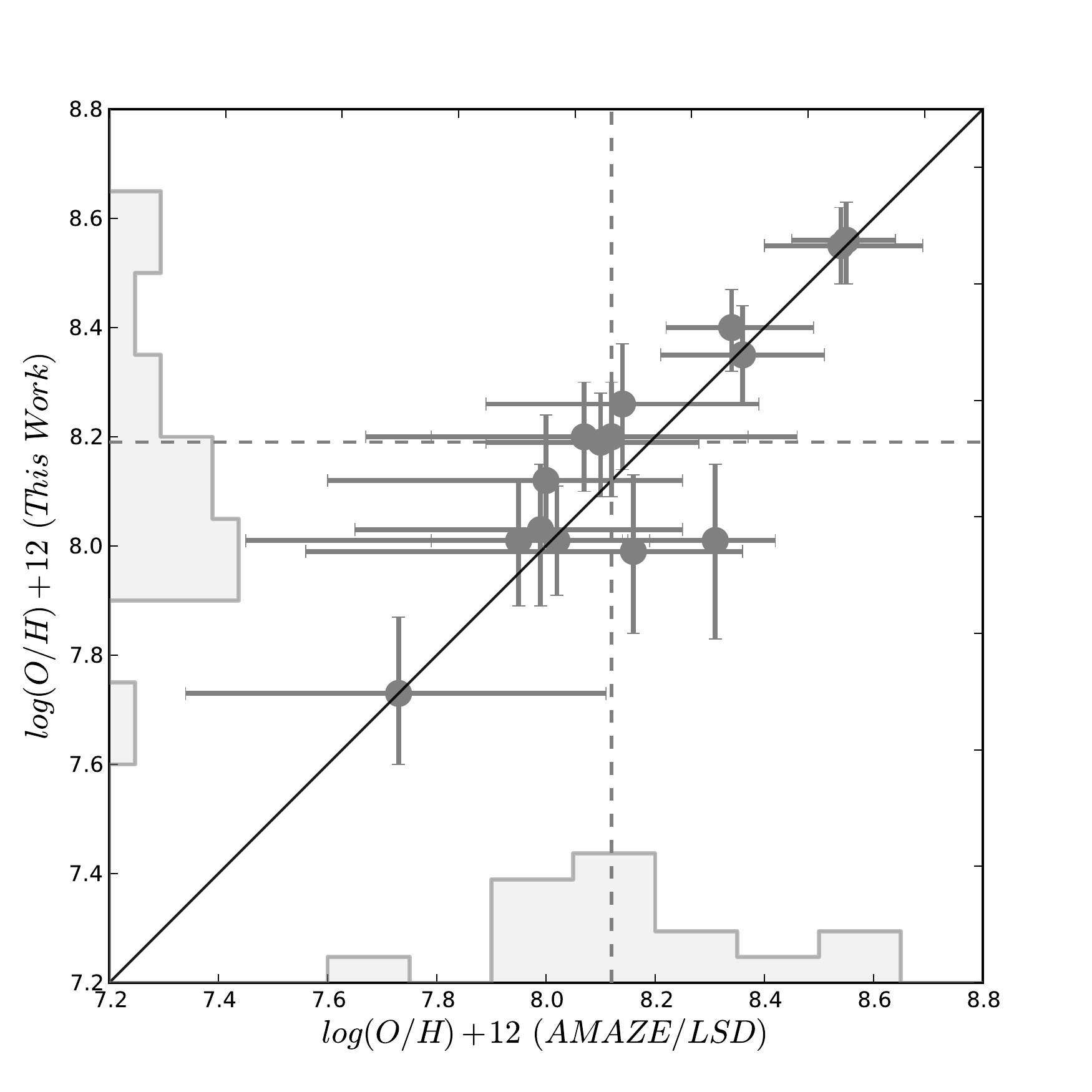}}
	\caption{A comparison of our measurement of metallicity for the z $\sim$ 3 galaxies in \citet{maiolino08} and \citet{mannucci09} compared to the metallicities quoted in those papers. The mean and median offset of the measured metallicities are 0.013 and 0.004 dex respectively. On each axis the histograms represent the distribution of the individual points and the dashed lines represent the median of those distributions.}
	\label{fig_z3_comprison}
	\end{figure}

To check for consistency we take the line ratios from the \citet{maiolino08} and \citet{mannucci09} papers, de-redden via the best-fitting extinctions quoted, and measure the metallicities in the same way as described above. We compare these measured metallicities to the metallicities quoted in those papers. Fig. \ref{fig_z3_comprison} shows the results of this test; our method agrees very well across the all 15 data points with median and average difference of 0.004 and 0.013 dex respectively.

%%%%%%%%%%%%%%%%%%%%%%%% PHOTOMETRY %%%%%%%%%%%%%%%%%%%%%%%%%%%%%%%%%%%%%%%

\section{Results: MZR and FMR}\label{results}

In this section we present the results of our study of the MZR and FMR, exploring the position of the 3D-$HST$ z$\gtrsim$2 emission line galaxies in the MZ plane and on the FMR surface. Throughout we compare to the previous z $\gtrsim$ 2 sample of \citet{erb06b} as it represents the largest single sample of z $\gtrsim$ 2 galaxy metallicities to date, and directly follows our method of stacking low S/N galaxies into bins of stellar mass. We also compare in Section \ref{sect_fmr} to the complimentary study of \citet{henry13b} at lower redshift (z $\sim$ 1.8) with metallicities measured from stacked grism spectra using the same nebular emission lines.

\subsection{Mass-Metallicity Relation}

\begin{figure}
	\centerline{\includegraphics[width=\columnwidth]{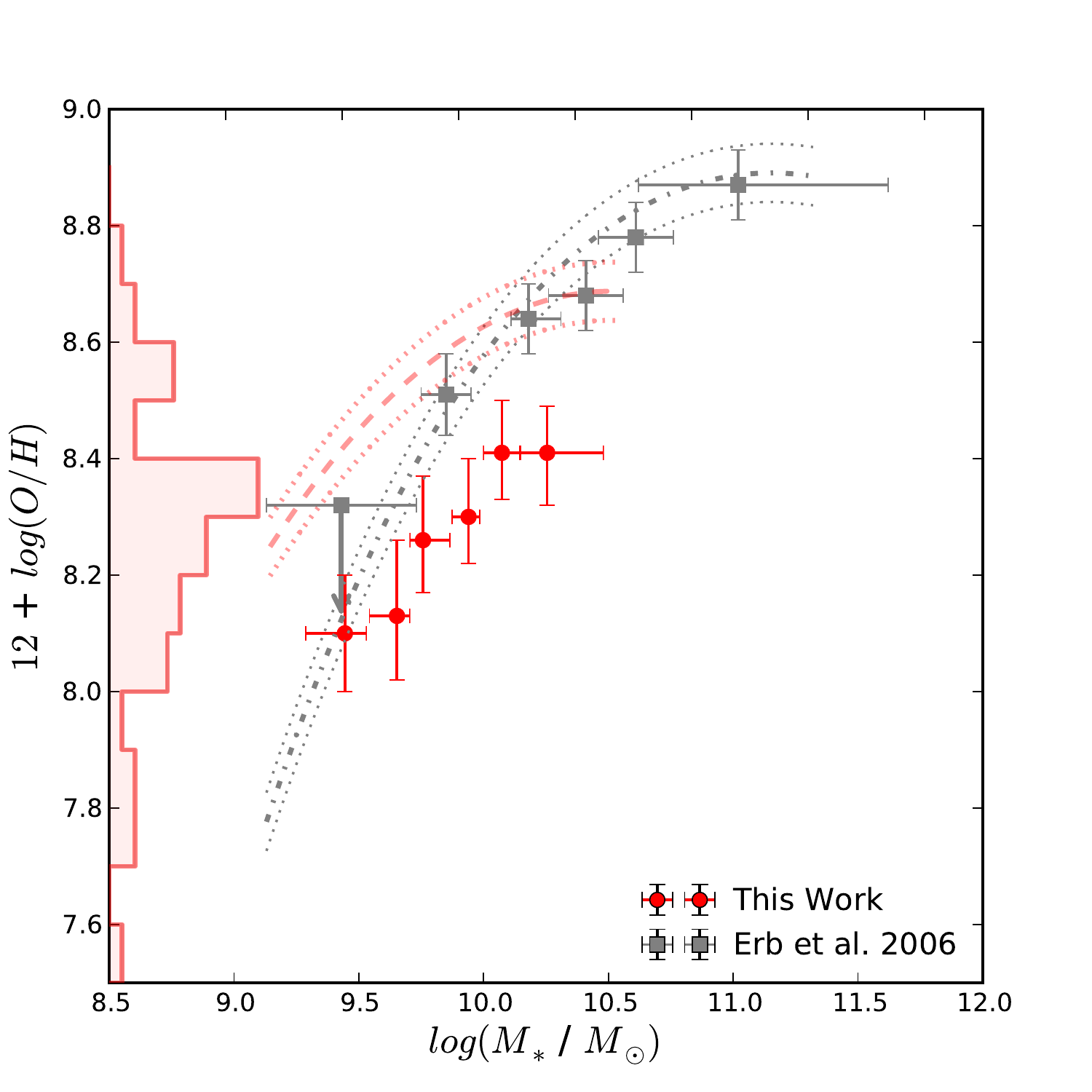}}
	\caption{The MZR for the z $\gtrsim$ 2 galaxies in our sample. The red circles represent the galaxy stacks presented in this paper with metallicities derived from the \oii, \oiii \ and \hb \ nebular emission lines. The red histogram shows the distribution of metallicities for the individual galaxies in the stacks with a measured metallicity. The grey squares represent the z $\gtrsim$ 2 MZR from \citet{erb06b} with metallicities derived from [NII] and \ha. The red dashed line (our sample) and the grey dot-dashed line \citep{erb06b} show the predicted positions of the two data sets in the M-Z plane derived from the FMR (see Sec. \ref{sect_fmr}).}
	\label{fig_MZ}
	\end{figure}

Fig. \ref{fig_MZ} shows the MZR for our data compared to the MZR from \citet{erb06b}. \citet{erb06b} measured metallicities from the \nii/\ha \ ratio using the calibration of \citet{pettini04}, so we convert their data to the \citet{maiolino08} calibration to keep the metallicity scales consistent. It is important that strong line metallicity diagnostics are compared using the same calibration since it has been noted that mismatches between different calibration scales can cause systematic offsets in derived metallicities \citep[e.g][]{kewley08}. \citet{erb06b} used a Chabrier IMF to derive stellar masses so that no mass conversion is necessary for consistency with our analysis. As a cross-check, we have run the photometric data from their paper through \textsc{lephare} and confirmed that we derive similar stellar masses.

Our data support the existence of the MZR at z $\sim$ 2. It can be seen from Fig. \ref{fig_MZ} that we find, in agreement with many other studies of the MZR, a decrease in metallicity with decreasing stellar mass. Across the range of stellar mass 9.3 $<$ log(M/M$_{\odot}$) $<$ 10.5 we observe a decrease of $\sim$ 0.3 dex in metallicity. Our data extends the \citet{erb06b} study as we probe masses below log(M/M$_{\odot}$) $\sim$ 10.0 where their data could only place an upper limit on the metallicity. However at a given value of stellar mass we measure lower metallicities than \citet{erb06b}. This discrepancy is of the order of $\sim$ 0.2 - 0.3 dex in the highest mass bins. This offset is large considering the similar redshifts of the two samples, in Section \ref{sect_fmr} we explore a possible explanation for this offset given by the FMR.

	\begin{figure}
	\centerline{\includegraphics[width=\columnwidth]{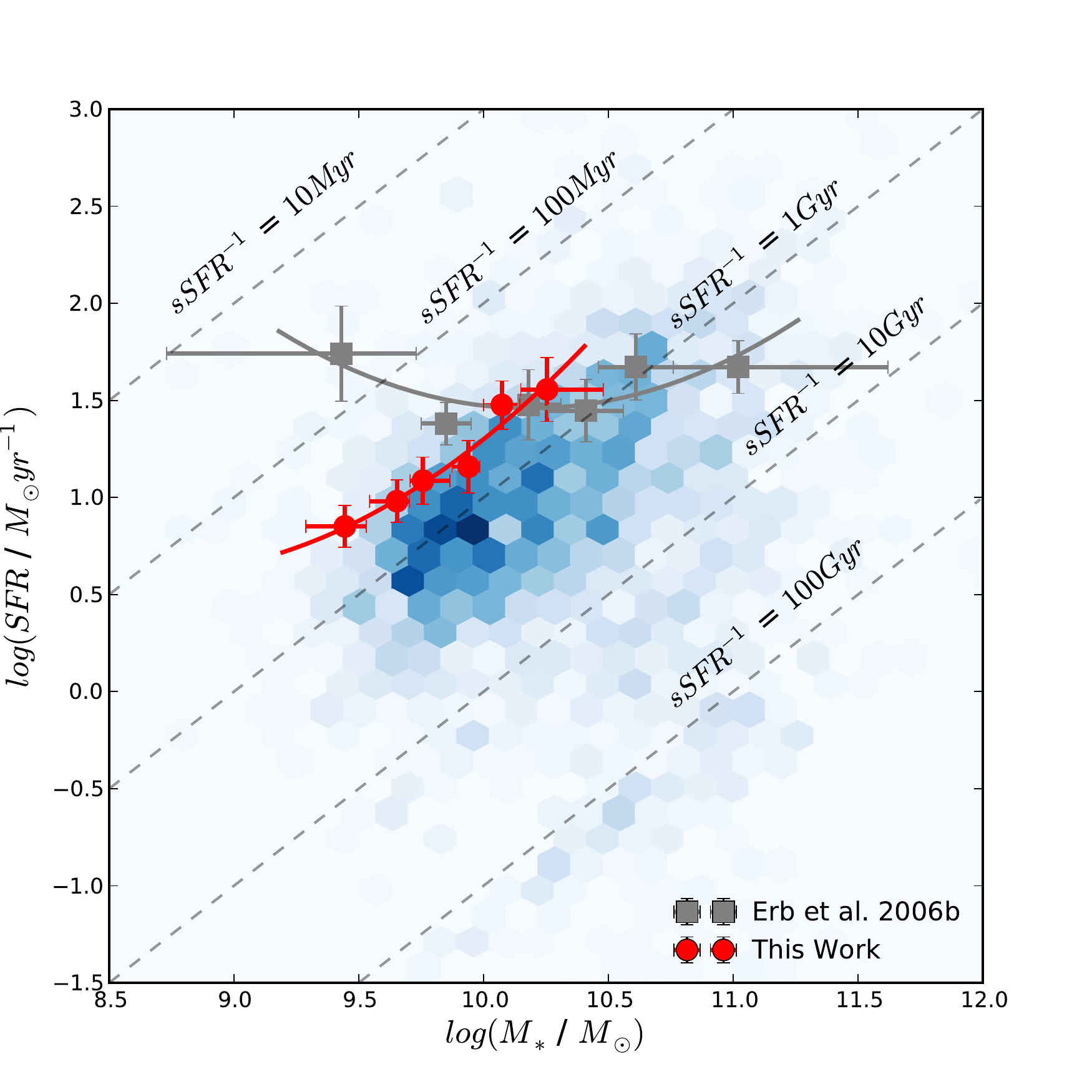}}
	\caption{The position of the galaxies in our sample in the M$_*$-SFR plane. Red circles represent SFRs for the stacked galaxies presented in this paper with SFR measured from the H$\beta$ flux of the stacked spectra, converted to H$\alpha$ via the common conversion factor H$\alpha$/H$\beta$ = 2.86 \citep{kennicutt98}. The grey squares represent SFRs for the stacked galaxies from \citet{erb06b} (grey squares) with SFR measured directly from H$\alpha$. The points are overlaid on a 2D histogram (in blue) of $\sim$ 3000 galaxies at 2 $<$ z $<$ 2.5 from \citet{whitaker2012} to provide a visual reference for the region of the M$_*$-SFR plane we are sampling. Polynomial fits to the \citet{erb06b} data (grey line) and the data presented here (red line) are indicated on the plot, these lines represent two slices along the FMR surface.}
	\label{fig_masssfr}
	\end{figure}

\subsection{Fundamental Metallicity Relation}\label{sect_fmr}

To investigate the FMR the SFR of the galaxy stacks were measured from the \hb \ line flux as described in Sec. \ref{sfr_hb}. We use the original FMR equation given in \citet{mannucci10} as opposed to its extension in \citet{mannucci11} as we find that the \citet{mannucci10} FMR was derived using a sample that best matches the mass-SFR parameter space of our sample and the \citet{erb06b} sample. The \citet{mannucci11} FMR extension was derived by adding low-mass (log(M / M$_{\odot}$) $<$ 9.2), low-SFR galaxies which are not representative of the galaxies in our data set. We find that the using \citet{mannucci11} FMR parametrization produces discontinuities in the M-Z relation at the high masses and high SFRs of our sample. However, we note that adopting the \citet{mannucci11} FMR parametrization does not significantly change the results discussed below.

According to the FMR, measuring lower metallicity in a given stellar mass bin should indicate that the average SFR in that bin is higher. Therefore, given that we measure lower metallicities than \citet{erb06b} across our stellar mass bins (Fig. \ref{fig_MZ}), we should also observe elevated average SFRs in those bins. However as illustrated in Fig. \ref{fig_masssfr} the average SFR within our stellar mass bins is lower than the \citet{erb06b} data, converging to similar values towards the higher mass bins in our sample ($\sim$ 10$^{10}M_{\odot}$). This implies at least one of these data sets is in contradiction to the FMR. The problem is not alleviated by taking our SFRs from SED fitting as these are on average slightly lower than the \hb \ derived SFRs as discussed in Sec. \ref{sfr_hb}. 

\begin{figure}
	\centerline{\includegraphics[width=\columnwidth]{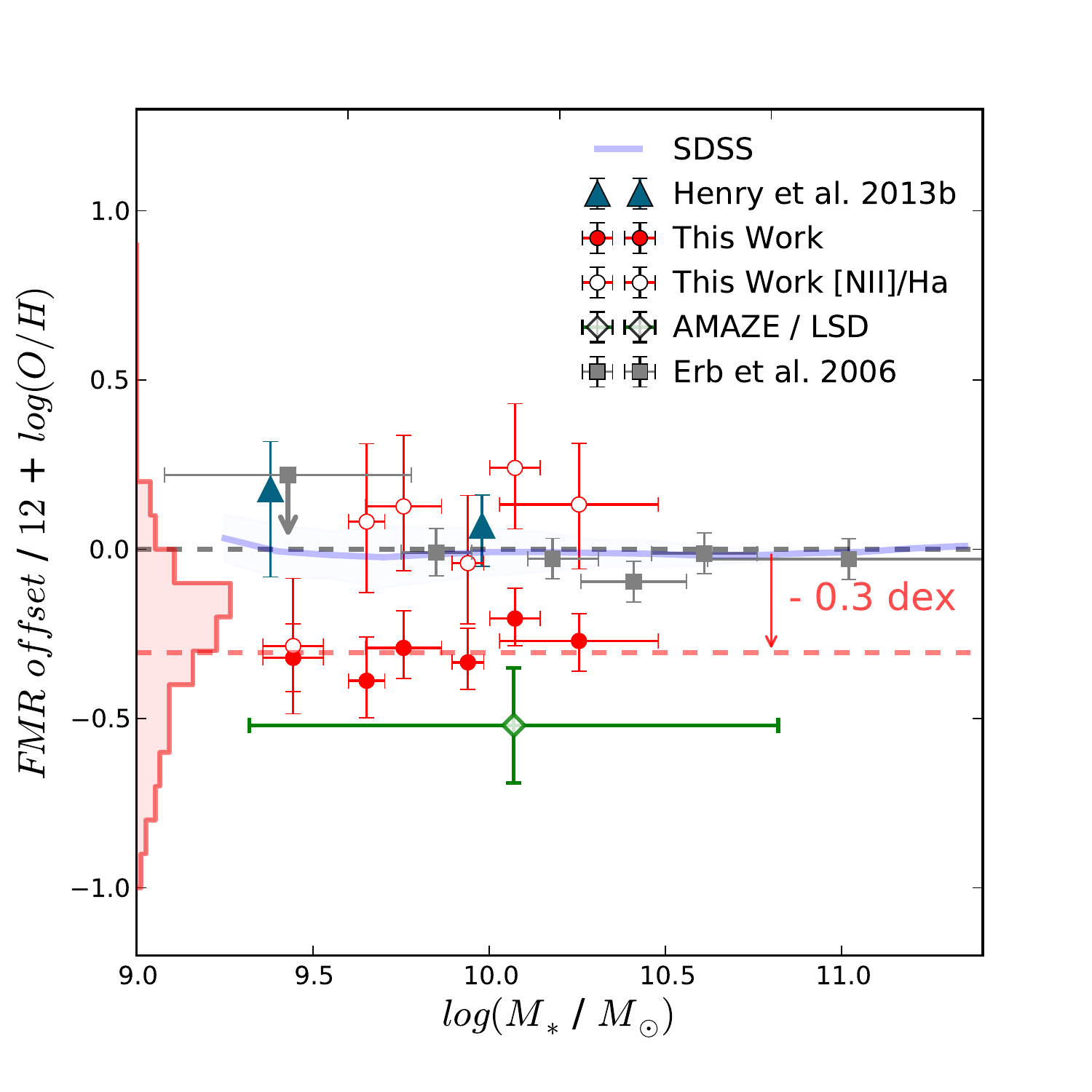}}
	\caption{The difference in the observed values of metallicity from those predicted by the FMR as a function of stellar mass for various samples. The blue line is the original SDSS sample of \citet{mannucci10} with the blue shaded region representing the 1$\sigma$ dispersion. The filled red circles represent the galaxy stacks presented in this paper. The green diamond shows a combination of z $\sim$ 3 galaxies from the AMAZE/LSD surveys \citep{maiolino08,mannucci09} and the grey squares show the \citet{erb06b} galaxies. The navy blue triangles show data from \citet{henry13b} for the mass bins in their data which overlap with our sample. Our results, within the error bars, are not consistent with the FMR derived in the local Universe persisting out to z $\gtrsim$ 2, and yield a average offset of $\sim$ 0.3 dex. The open red circles represent the FMR offset predicted from the \nii/\ha \ ratio of our sample using theoretical predictions for the evolution of ionization conditions in star-forming regions with redshift (see Sec. \ref{sec_photoion} for discussion)}
	\label{fig_fmr_residual}
	\end{figure}

The FMR offset of various samples is illustrated in Fig. \ref{fig_fmr_residual}. This figure shows the difference between the metallicity observed and the metallicity predicted from the FMR given the \hb \ derived SFR and the median values of stellar mass for each galaxy stack. It can be seen that the galaxies presented in this paper lie offset from the FMR by an average of $\sim$ 0.3 dex (dashed red line) whilst the \citet{erb06b} data are consistent with the FMR. This is also illustrated in Fig. \ref{fig_MZ} where the predicted positions of the two samples, based on the FMR, are shown in the MZ plane.

Also shown in Fig. \ref{fig_fmr_residual} is the  $\sim$ 0.5 dex offset of the AMAZE/LSD galaxies \citep{maiolino08,mannucci09} at z $\sim$ 3. The offset was calculated by taking the median values of mass, metallicity, and SFR of the individual galaxies quoted in those papers. We convert the stellar mass from \citet{maiolino08} to be consistent with the Chabrier IMF used in this analysis. This offset of the z $\sim$ 3 galaxies was also noted in \citet{mannucci10}. As discussed in Section \ref{data_metallicity} we use the same calibration and the same set of emission lines to measure metallicities as the AMAZE/LSD surveys. Furthermore, our method of metallicity measurement is shown to return consistent results (see Fig. \ref{fig_z3_comprison}).

The fact that the z $\gtrsim$ 2 galaxies presented here are offset from the FMR, as are the z $\sim$ 3 galaxies from AMAZE/LSD, whilst the z $\gtrsim$ 2 galaxies of \citet{erb06b} are in agreement, suggests that the choice of metallicity indicator may be affecting the measured metallicities. At first sight it appears that this should not be the case, since all line ratios are calibrated to the same metallicity scale. However, these calibrations were made using local Universe star-forming galaxies which may not be representative of typical star-forming galaxies at high-redshift. In Sec. \ref{sec_photoion} we investigate whether a change in the ionization conditions of HII regions at high redshifts affects the consistency of metallicity measurements using different sets of line ratios.

We also compare our results with data from \citet{henry13b} in Fig. \ref{fig_fmr_residual} (blue triangles). The \citet{henry13b} sample consists of 83 grism spectra in the redshift range 1.3 $<$ z $<$ 2.3, the spectra are stacked in four bins of stellar mass and metallicities are measured from the oxygen and \hb \ lines using the \citet{maiolino08} calibrations. In this respect our data and methods are very similar. The main difference between our data sets is that their sample spans a much wider redshift range resulting in a lower median redshift (z = 1.76), and the \citet{henry13b} sample probes to lower stellar masses (log(M/M$_{\odot}$) $<$ 10$^{9.2}$). In Fig. \ref{fig_fmr_residual} we only include the mass bins which overlap with our sample. Despite using the same line diagnostics the data from \citet{henry13b} lie on the FMR relation, in disagreement with our results. However, their results are obtained by stacking over a wide redshift range, with a median redshift of (z = 1.76), which is significantly lower than ours (z = 2.16), again an evolution in ionization conditions may explain the inconsistency as we discuss in Sec. \ref{sec_photoion}.

%%%%%%%%%%%%%%%%%%%% RESULTS %% %%%%%%%%%%%%%%%

\subsection{Photoionization Conditions}\label{sec_photoion}

The metallicity calibrations outlined in \citet{maiolino08} are calibrated against local star-forming galaxies. The line flux ratios used in these calibrations are known to have other dependencies (e.g. ionization parameter and gas density) and therefore to use them at high redshift we have to assume that the conditions in star-forming galaxies at high redshift do not differ significantly from those at low redshift. However, it is known that many properties of galaxies change at high redshift; for example at z $\sim$ 2, galaxies are on average more compact and have higher SFR \citep[e.g.][]{buitrago2008, mclure12, whitaker2012}. These changes may affect the physical conditions within these galaxies. Below, we discuss the evolution of physical conditions in HII regions via the BPT diagram.

\subsubsection{BPT diagram}

\begin{figure}
	\centerline{\includegraphics[width=\columnwidth]{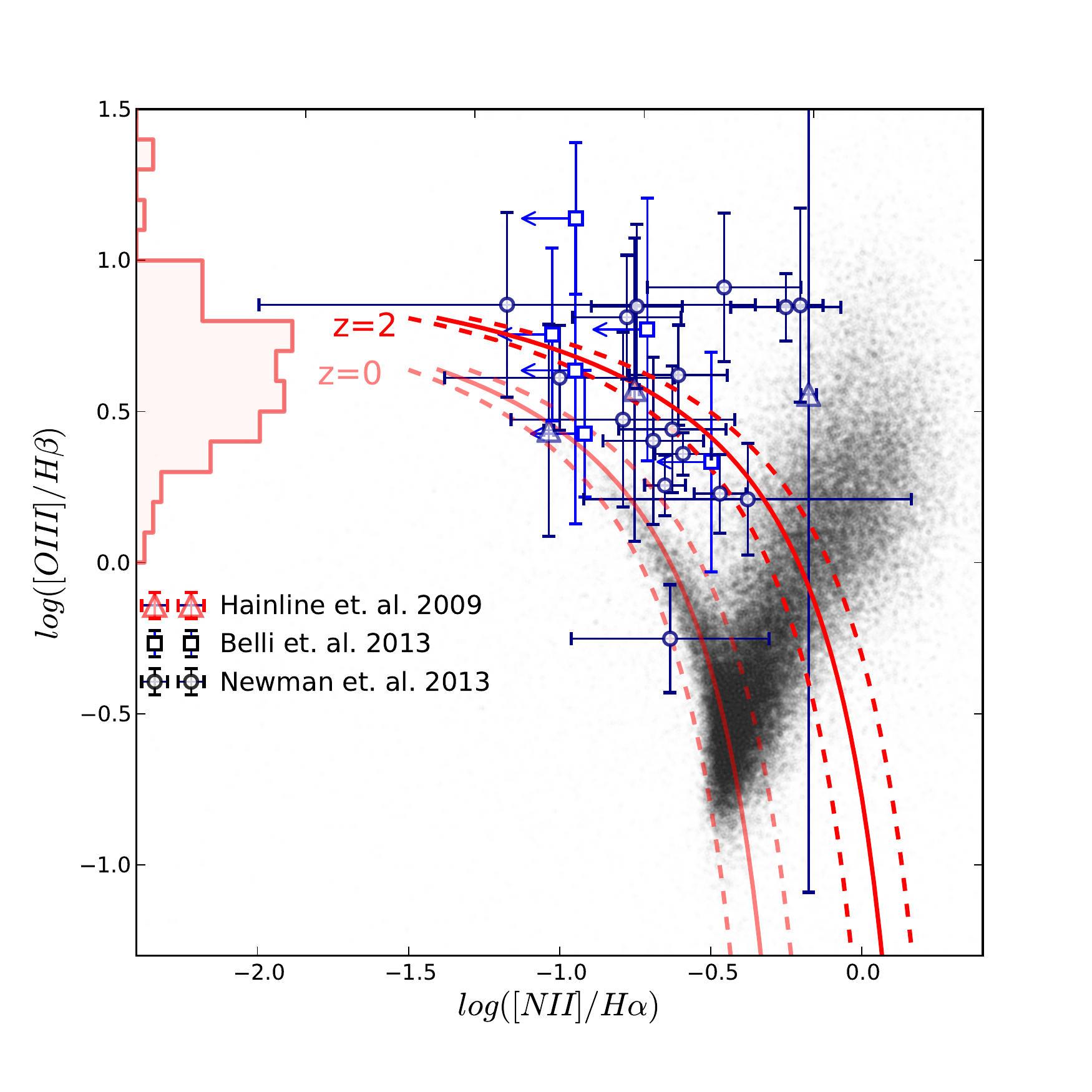}}
	\caption{The BPT diagram of log([NII]/H$\alpha$) versus. log([OIII]/H$\beta$) used for diagnosing HII regions via emission-line ratios. The grey points show a sample of local SDSS galaxies from \citet{brinchmann13}. The blue data points with error bars show the current data for z $\sim$ 2 galaxies with measurements in all four emission lines, the open triangles show three galaxies from \citet{hainline09}, the open squares show a sample of gravitationally lensed galaxies from \citet{belli13} and the open circles are from \citet{newman13}. A histogram of the distribution of the \oiiib/\hb \ ratio of the individual galaxies in our sample with measured \hb \ is shown on the y axis	. The two red lines show the star-forming sequence of galaxies in the BPT diagram at z = 0 and z = 2 respectively, taken from the theoretical calibration of \citet{kewley13a}.}
	\label{fig_bpt}
	\end{figure}

The most commonly used diagnostic of photoionization conditions in HII regions is the BPT diagram \citep*{baldwin81} which has traditionally been used to separate star-forming galaxies from AGN in local galaxies. In recent years the BPT diagram has been used to suggest that the ionization conditions of HII regions at z $>$ 1 are different from those in the local Universe \citep[e.g.][]{shapley05, erb06a, liu08, hainline09, nakajima13}.  The BPT diagram is shown in Fig. \ref{fig_bpt}. Plotted in black are a sample of galaxies from the SDSS DR7 MPA-JHU data release \citep{brinchmann13} in the redshift range 0.04$<$z$<$ 0.5. Fig. \ref{fig_bpt} also shows a handful of galaxies at z$\sim$2  which have \oiiib, \hb, \ha \ and \nii \ measurements and can therefore be placed in the BPT diagram \citep{hainline09, belli13,newman13}. It is evident that z $\sim$ 2 galaxies are offset from the local BPT relation for star-forming galaxies, particularly striking are the large values of the \oiiib /\hb \ ratio seen at these redshifts, clearly offset from the locus of the SDSS galaxies. The offset is also seen at lower redshifts, for example in the DEEP2 survey of 1 $<$ z $<$ 1.5 star-forming galaxies of \citet{shapley05} and the \citet{trump13} study of z $\sim$ 1.5 galaxies using MOSFIRE and 3D-$HST$. For our sample, since we can only measure the \oiiib /\hb \ ratio, we show a histogram of the values for our full galaxy sample on the y axis of Fig. \ref{fig_bpt}. Our results are consistent with previously measured z $\sim$ 2 ratios of individual objects, the median \oiiib/\hb \ ratio of our sample is 0.59. Of course, elevated values of the \oiiib /\hb \ ratio are not necessarily an indication of change in ionization conditions in HII regions as the ratio is also strongly metallicity dependent.

However, \citet{brinchmann08} and \citet{kewley13a,kewley13b} have investigated the evolution of the star-forming sequence with redshift and claim that a larger ionization parameter, higher electron density and harder ionizing radiation field may be the cause of this evolution. \citet{trump13} on the other hand claim to detect nuclear activity in 2/3 of a sample of z $\sim$ 1.5 galaxies, arguing that the majority of high-redshift galaxies show evidence for nuclear activity, such that high-redshift galaxies are offset from the star-forming line in the BPT diagram in the same way as AGN are offset in local galaxies. However it is not clear that AGN diagnostics for local Universe galaxies can be directly applied to galaxies at high redshift given a change in ionization conditions \citep{kewley13b}. Most recently \citet{kewley13a} used cosmological hydrodynamical simulations to investigate how the positions of star-forming galaxies in the BPT diagram evolve given different sets of assumptions for the ionization conditions within galaxies, concluding that the position changes depending on the hardness of ionizing radiation field, ionization parameter, and electron densities. \citet{kewley13a} provide a redshift dependent equation giving the position for the star-forming sequence of galaxies in the BPT diagram, the position of this main sequence at z = 0 and z = 2 is shown in Fig. \ref{fig_bpt}. These star-forming sequences can be seen to be consistent with both the SDSS and current z $\sim$ 2 data.

We note that the z $\sim$ 2 star-forming sequence taken from \citet{kewley13a} is derived using the upper limit of the hardness of the ionizing radiation field, and therefore represents the most extreme star-forming conditions realised in their models at this redshift (the z = 0 sequence being the lower limit at any redshift). As noted in \citet{kewley13a}, until more statistically significant samples of galaxies become available at high redshift it is only possible to assume an upper and lower limit of the star-forming sequence at any given redshift.

\subsubsection{O32 versus. R$_{23}$ diagram}

\begin{figure}
	\centerline{\includegraphics[width=\columnwidth]{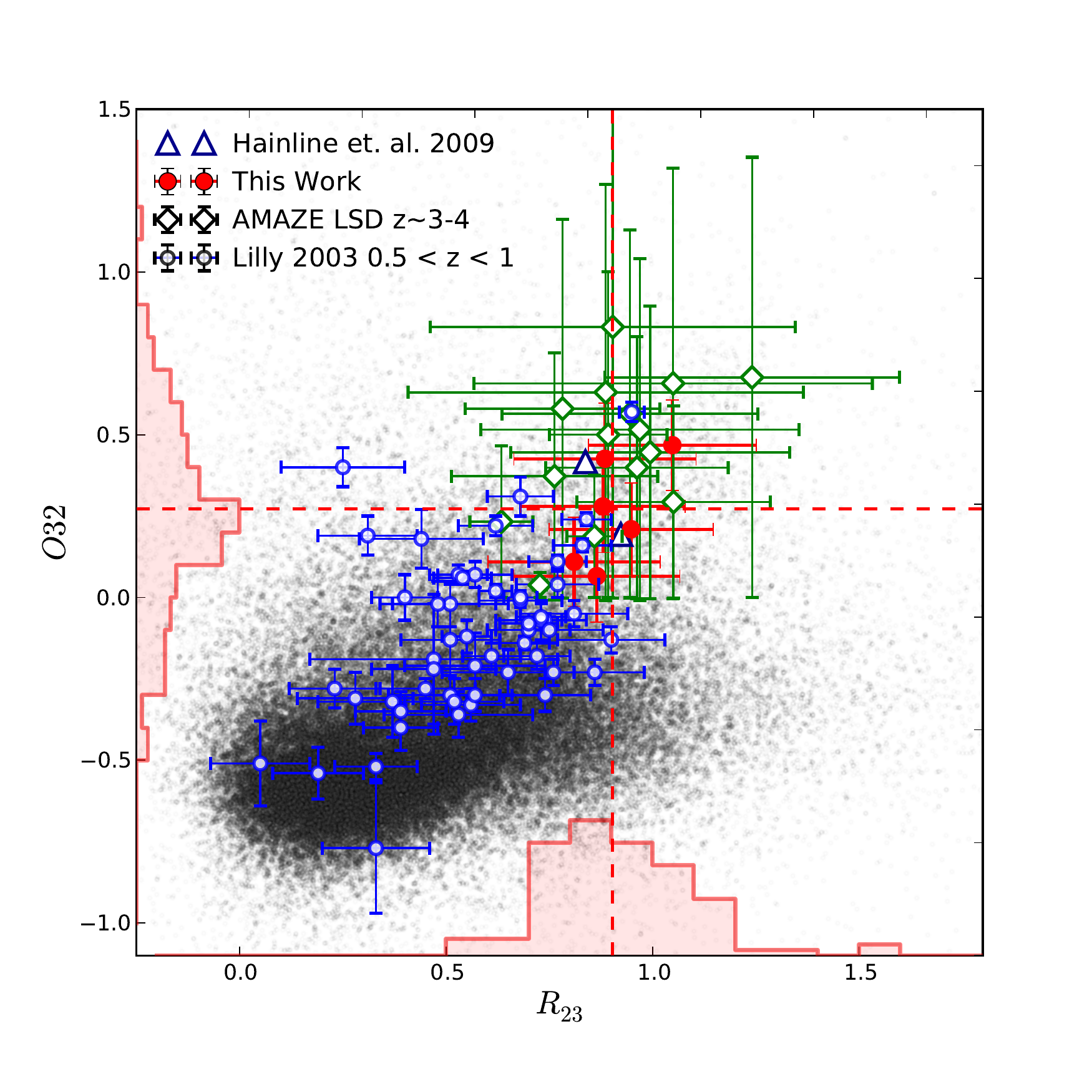}}
	\caption{The O32 versus R$_{23}$ diagnostic diagram used to differentiate change in ionization conditions from a change in metallicity of galaxies \citep{lilly03, hainline09,nakajima13}. The stacked galaxies in our sample are represented by filled red circles and the distribution of the individual galaxies are shown as histograms on each axis. The open blue circles represent a sample of 0.5 $<$ z $<$ 1 galaxies taken from \citet{lilly03}, the open green diamonds are the AMAZE/LSD sample of z $\sim$ 3 -4 galaxies from \citet{maiolino08} and \citet{mannucci09}.}
	\label{fig_o32}
	\end{figure}

We can investigate the ionization state of our z $\gtrsim$ 2 galaxies directly using the O32 versus. R$_{23}$ diagram \citep{lilly03}. The \oiii/\oii \ ratio (O32) is ionization-parameter sensitive \citep{kewley02, brinchmann08} and we can use this to test the typical ionization conditions in our galaxy sample. However, O32 is also sensitive to metallicity \citep{nagao06}, so a second more metallicity sensitive ratio must be used to break the degeneracy. We follow the method employed in various studies in the literature by plotting O32 versus R$_{23}$ \citep[e.g.][]{lilly03,hainline09,nakajima13}. The comparison is shown in Fig. \ref{fig_o32} where we again plot the SDSS sample plus a sample at intermediate redshift (0.47 $<$ z $<$ 0.92) from \citet{lilly03}. Our data seem to support the conclusions of \citet{hainline09} and \citet{nakajima13} that galaxies at z $>$ 2 are systematically offset toward larger values of O32 at a given value of R$_{23}$. As discussed by those authors this is evidence for higher ionization parameters at fixed metallicity. Therefore, as our data are consistent with an elevated ionization parameter relative to local galaxies, indicative of a harder ionization radiation field at z $\sim$ 2, the galaxies in our sample should lie offset from the z = 0 star-forming galaxy sequence, and on, or towards, the z = 2 star-forming sequence as shown in Fig. \ref{fig_bpt}.

\subsubsection{Implications for metallicity measurements}\label{metal_implications}

\begin{figure}
	\centerline{\includegraphics[width=\columnwidth]{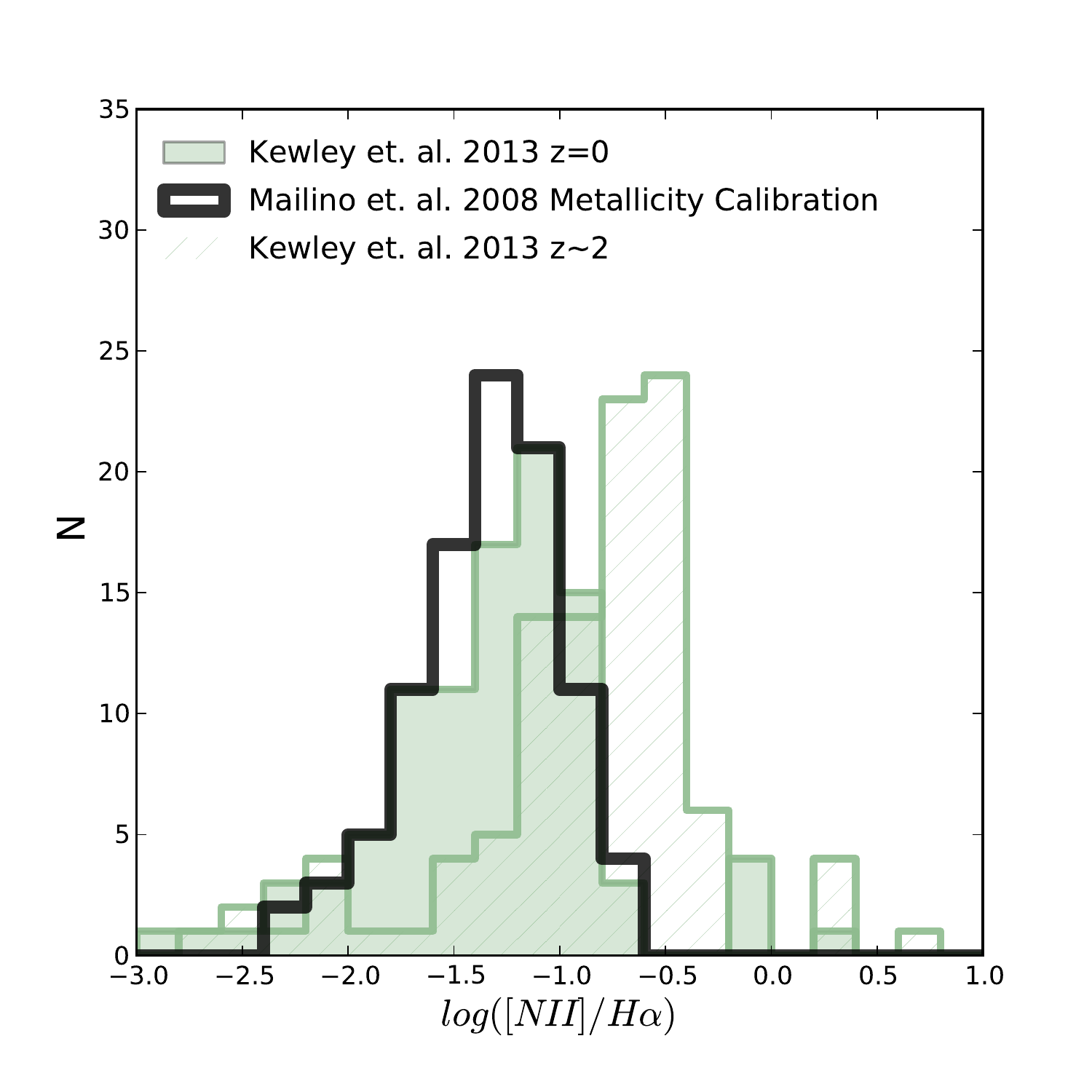}}
	\caption{Histograms showing the inferred \nii/\ha \ ratios for the individual galaxies in our sample with detections in all lines. The open black histogram is the \nii/\ha \ ratio derived using the measured metallicity with the \citet{maiolino08} metallicity calibrations (Eq. \ref{eq:m08niiha}). The filled green histogram is the ratio derived from the measured \oiiib/\hb \ ratio using the BPT diagram star-forming galaxy main sequence equation at z=0 from \citet{kewley13a} (Eq. \ref{eq:O3_N2_seq_z}). The median and median absolute deviation (MAD) of the ratios derived from \citet{maiolino08} and \citet{kewley13a} (z = 0) respectively are -1.34 $\pm$ 0.2 dex and -1.26 $\pm$ 0.3 dex. The hatched green histogram is derived in the same way as the filled green histogram except using the galaxies redshift in Eq. \ref{eq:O3_N2_seq_z} (z $\sim$ 2). The median and MAD of this distribution is -0.74 $\pm$ 0.3 dex.}
	\label{fig_nii_ha_hist}
	\end{figure}

An offset in the position of star-forming galaxies in the BPT diagram has implications for the metallicities derived from either the oxygen and \hb \ or \nii \ and \ha \ emission-line ratios. Since the metallicity calibrations of \citet{maiolino08} were made using local star-forming galaxies, metallicities derived from either oxygen and \hb \ lines, or \nii/\ha, will only be consistent for galaxies which lie on the z = 0 star-forming sequence. Fig. \ref{fig_nii_ha_hist} illustrates this by comparing the \nii / \ha \ ratios inferred in three different ways for the individual galaxies in our sample with detections in all emission lines. First, we use the measured metallicities from oxygen and \hb \ lines and work backwards from the \citet{maiolino08} calibrations to infer a \nii/\ha \ ratio via:

\begin{equation}
\label{eq:m08niiha}
\begin{array}{rl}
log([NII]/H\alpha)=&-0.7732+1.2357x-0.2811x^2\\
				   &-0.7201x^3-0.3330x^4\\
\end{array}
\end{equation}
Where $x$ = 12+log(O/H) - 8.69. 

Secondly and thirdly, we take the measured values of the \oiiib/\hb \ ratio and use the BPT star-forming sequence of \citet{kewley13a} at z $\sim$ 2 and 0, respectively, to infer a \nii/\ha \ ratio via:

\begin{eqnarray}
\log({\rm [NII]/H}\alpha) & = & 0.1833z - 0.08    \nonumber \\
& & + \frac{0.61}{\log({\rm [OIII]/H}\beta) - 1.1 - 0.03z}  \label{eq:O3_N2_seq_z}.
\end{eqnarray}
Though we note, as described in Sec. \ref{sec_photoion}, that at z $\sim$ 2 this represents an upper limit on the \nii/\ha \ ratio assuming extreme star-formation conditions

It can be seen from Fig. \ref{fig_nii_ha_hist} that assuming the \citet{maiolino08} calibrations leads to systematically lower predicted values of the \nii/\ha \ ratio compared to those derived using the \citet{kewley13a} star-forming sequence at z $\sim$ 2. By contrast the \citet{maiolino08} calibrations are in good agreement with \nii/\ha \ ratios derived using \citet{kewley13a} star-forming sequence at z = 0. This implies the \citet{maiolino08} calibrations may not be applicable to high-redshift galaxies since, given a metallicity measured from the oxygen and \hb \ lines, they predict \nii/\ha \ ratios consistent with the local star-forming galaxy sequence, whereas z $\sim$ 2 galaxies lie offset from this sequence towards higher \nii/\ha \ ratios.

The effect of assuming the \citet{kewley13a} star-forming sequence on the FMR is shown in Fig. \ref{fig_fmr_residual} where the open red circles represent the metallicities derived from the \nii/\ha \ ratio assuming the \citet{kewley13a} conversion at the median redshift of the stacks (z $\sim$ 2).Metallicities are then derived from the \nii/\ha \ ratio using the \citet{maiolino08} calibrations. These theoretical \nii/\ha \ ratios return systematically higher metallicities than those derived from oxygen and \hb \ lines. This suggests that metallicities at high redshift based on the oxygen and \hb \ lines are not directly comparable to metallicities measured from the \nii/\ha \ ratio using local empirical calibrations. It can also be seen from Fig. \ref{fig_fmr_residual} that these theoretical \nii/\ha \ metallicities are in better agreement with the FMR. Thus, the consistency of the \citet{erb06b} data with the FMR, and the inconsistency of our data with it, may be explained by this evolution of galaxies in the BPT diagram with redshift.

The evolution in ionization conditions of galaxies with redshift may also explain the apparent discrepancy between our data and the \citet{henry13b} results. Since their sample spans a much larger range in redshifts (1.3 $<$ z $<$ 2.3), and therefore a larger range of ionization conditions at a given metallicity, it is perhaps not surprising that in a given mass bin our metallicities are offset. However a detailed analysis of the effect of stacking across a large range in redshift is beyond the scope of this work, and we focus here on the discrepancy between \nii/\ha \ and oxygen and \hb \ metallicities using local empirical calibrations at high redshift.

\subsubsection{\citet{newman13} data}

To further investigate the discrepancy between line indicators, we took a sample of 11 z $\sim$ 2 galaxies from \citet{newman13} which have flux measurements in the \nii, \ha, \oiiib \ and \hb \ lines, and have measured masses and SFRs. The masses were derived using a \citet{chabrier03} IMF, so no conversion was necessary for comparison with our results. Metallicities were derived from the \nii/\ha \ and \oiiib/\hb \ line ratios individually and the offset from the predicted FMR value computed as described in Sec. \ref{sect_fmr}. 

\begin{figure}
	\centerline{\includegraphics[width=\columnwidth]{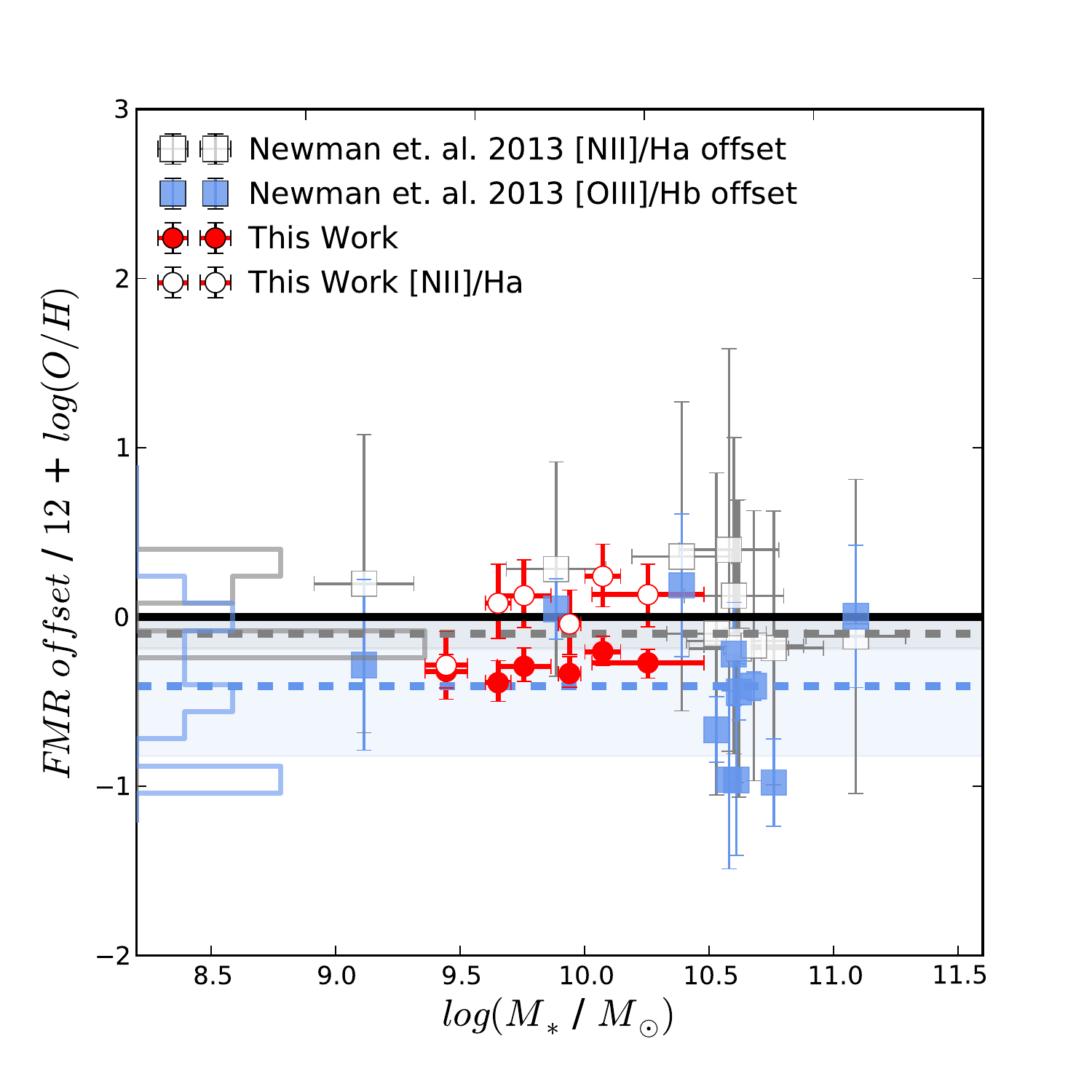}}
	\caption{The FMR offset for a sample of z $\sim$ 2 star-forming galaxies taken from \citet{newman13}. Open grey squares show the metallicities derived from the \nii/\ha \ ratio and the filled blue squares show the metallicities derived from the \oiiib/\hb \ ratio. On the y axis the grey and blue histograms show the distribution of the individual points with the horizontal dashed lines representing the median values of the distributions. The median and MAD for the \nii/\ha \ ratio and \oiiib/\hb \ ratio respectively are -0.10 $\pm$ 0.09 dex and -0.44 $\pm$ 0.22 dex., the MAD is shown by the shaded region around the median line. For reference the galaxies presented in this paper are plotted as filled red and open red circles as in Fig. \ref{fig_fmr_residual}.}
	\label{fig_newman13}
	\end{figure}

Note that the full set of oxygen and \hb \ calibrations used on our data could not be used on the \citet{newman13} data so oxygen and \hb \ metallicities were derived from the \oiiib/\hb \ ratio alone. When used independently the \oiiib/\hb \ metallicity calibration is double-valued (see \citet{maiolino08} and Fig. \ref{fig_metallicity_curves}), so to derive a metallicity from the \oiiib/\hb \ ratio solely it is necessary to choose between the upper and lower solutions. For each galaxy in the \citet{newman13} sample the upper solution was taken, as it was in agreement with the high metallicities implied from the \nii/\ha \ solution. Nevertheless, even when taking the upper solution for \oiiib/\hb, the \nii/\ha \ ratio returns systematically higher values of metallicity in better agreement with the FMR. This is illustrated in Fig. \ref{fig_newman13}. The median and MAD for the \nii/\ha \ ratio and \oiiib/\hb \ ratio respectively are -0.10 $\pm$ 0.09 dex and -0.44 $\pm$ 0.22 dex. It is also interesting to note that the \citet{newman13} data probe a higher-mass regime, similar to the \citet{erb06b} sample, indicating this effect is seen across a wide range in mass $\sim$ 9.5 $<$ log(M/M$_{\odot}$) $<$ 11.5.

To summarize, these arguments suggest that if spectra are obtained in the K band for our galaxy sample such that the \nii/\ha \ ratio can be measured, this ratio will return a higher metallicities by $\sim$ 0.3 - 0.5 dex over metallicities derived from the oxygen and \hb \ line using the \citet{maiolino08} metallicity calibrations. Thus from the \nii/\ha \ ratio one will observe better consistency with the FMR as in \citet{erb06b}, whilst from the oxygen and \hb \ lines an offset will be observed as in this work and at z $\sim$ 3 \citep{maiolino08,mannucci09}. A detailed comparison of the two line ratios as metallicity indicators is beyond the scope of this work, though resolving this issue is clearly of key importance if we want to conclude anything more definitive regarding the FMR at high redshift.

%%%%%%%%%%%%%%%%%%%% SUMMARY AND CONCLUSION  %% %%%%%%%%%%%%%

\section{Summary and Conclusions}\label{discussion}

We have selected a sample of \ngal galaxies in the redshift range 2.0 $<$ z $<$ 2.3 from an independent reduction of the 3D-$HST$ spectroscopic grism survey. In this redshift range the \oii, \oiii \ and \hb \ emission lines fall within the wavelength range of the grism spectra. Our aim is to use those emission lines to measure the metallicities of galaxies and measure their masses from the ancillary CANDELS photometry available in the 3D-$HST$ survey fields. We stack the galaxies in bins of stellar mass and construct a mass-metallicity relationship for this sample, which we can directly compare with the previous z $\gtrsim$ 2 study of \citet{erb06b}. We then measure the SFR from the \hb \ line and use this to investigate the FMR at these redshifts. Below is a summary of our results.

\begin{itemize}

  \item We measure the metallicities of our galaxies from the \oii, \hb  \ and \oiii \ emission lines via the calibrations of \citet{maiolino08}. We find a MZR in our galaxy sample consistent with the MZR reported elsewhere in the literature \citep[e.g.][]{erb06b, mannucci09, yuan13} in that observe a decrease in metallicity with a decrease in mass of our galaxy stacks (Fig. \ref{fig_MZ}). However our MZR is offset to lower metallicities at a given stellar mass from the z $\gtrsim$ 2 MZR of \citet{erb06b}. 

  \item We investigate this metallicity offset using the FMR proposed by \citet{mannucci10} which incorporates mass, metallicity and SFR in an attempt to explain the scatter and redshift evolution of the MZR. However we find our data are apparently inconsistent with the FMR. We measure metallicities lower by $\sim$ 0.3 dex from those predicted from the FMR given our measured masses and SFRs (Fig. \ref{fig_fmr_residual}). The previous z $\gtrsim$ 2 data of \citet{erb06b} are consistent with the FMR as discussed in \citet{mannucci10}, therefore there is a discrepancy between the current z $\gtrsim$ 2 data. One difference between the \citet{erb06b} study and our own is the metallicity indicator used. \citet{erb06b} use the \nii/\ha \ ratio whereas our metallicities are based on the oxygen and \hb \ lines. Interestingly our method for metallicity measurement follows previous z $\sim$ 3 MZR studies \citep{maiolino08, mannucci09} who find a similar offset from the FMR using the same set of emission lines.

  \item We investigate the ionization conditions of our galaxies to attempt to explain the FMR offset. We construct a O32 versus R$_{23}$ diagram following the method of \citet{lilly03}, \citet{hainline09} and \citet{nakajima13} (Fig \ref{fig_o32}). We find, consistent with previous z $\gtrsim$ 2 data, evidence for a enhancement of the O32 ratio at fixed value of R$_{23}$ indicative of an enhanced ionization parameter in these galaxies at fixed metallicity. 

  \item We note that from small samples of z $\sim$ 2 star-forming galaxies with \oii, \hb , \oiii , [NII] and H$\alpha$ measured \citep[e.g.][]{hainline09, belli13, newman13}, there is evidence that higher ionization parameters cause an offset from the star-forming galaxy sequence in the BPT diagram of local galaxies (Fig. \ref{fig_bpt}). This offset has recently been quantified by \citet{kewley13a} who provide an equation for the evolution of the star-forming sequence with redshift (Fig. \ref{fig_bpt} and Eq. \ref{eq:O3_N2_seq_z}). Given the evidence for an increased ionization parameter in our galaxies, and \oiiib/\hb \ ratios consistent with other z $\gtrsim$ 2 samples, we conclude our galaxies would most likely also lie offset from the z=0 star-forming galaxy sequence.

  \item Working backwards from our metallicity measurements we can infer the expected [NII]/H$\alpha$ ratio of our galaxies from the \citet{maiolino08} calibrations (Eq. \ref{eq:m08niiha}). We can also infer the expected \nii/\ha \ ratio from the \citet{kewley13a} BPT main sequence calibration at z $\sim$ 2 and z = 0 (Eq. \ref{eq:O3_N2_seq_z}). The inferred ratios from the \citet{maiolino08} calibrations are systematically lower than the inferred ratios from \citet{kewley13a} at z $\sim$ 2 but consistent with those at z = 0 (Fig. \ref{fig_nii_ha_hist}). This implies the \oiiib/\hb \ and \nii/\ha \ metallicity calibrations are not comparable for galaxies significantly offset from the star-forming main sequence of the BPT diagram.

  \item Taking the inferred \nii/\ha \ ratios from the \citet{kewley13a} calibration at z $\sim$ 2 we find our galaxies would fall into better agreement with the FMR (Fig. \ref{fig_fmr_residual}). For further investigation we also take a sample of 11 z $\sim$ 2 galaxies with measured \nii, \ha, \oiiib \ and \hb \ fluxes from \citet{newman13} and confirm that the metallicities derived using the \citet{maiolino08} calibrations are systematically higher than when using the \nii/\ha \ ratio (Fig. \ref{fig_newman13}), following the trend in our data inferred using \citet{kewley13a}. This further supports the conclusion that metallicities derived from the oxygen and \hb \ lines are not equivalent to those derived using \nii/\ha \ at high redshift.

  \item We conclude the discrepancy between our data and the FMR, as well as the z $\sim$ 3 data of \citet{maiolino08} and \citet{mannucci09}, may be the result of using locally calibrated empirical metallicity relations. If this is the case, the evolution of the FMR at high redshift cannot be quantified using these empirical calibrations which do not fully account for the change in ionization conditions in star-forming galaxies with redshift.

\end{itemize}

\section{Acknowledgments}
The authors would like to thank the anonymous referee for their useful comments that have helped improved the paper. FC and MC acknowledge the support of the Science and Technology Facilities Council (STFC) via the award of an STFC Studentship and an STFC Advanced Fellowship, respectively. RJM acknowledges the support of the European Research Council via the award of a Consolidator Grant (PI McLure). JSD acknowledges the support of the European Research Council via the award of an Advanced Grant, and the support of the Royal Society via a Wolfson Research Merit Award. This work is based on observations taken by the 3D-$HST$ Treasury Program (GO 12177 and 12328) with the NASA/ESA $HST$, which is operated by the Association of Universities for Research in Astronomy, Inc., under NASA contract NAS5-26555. This work is based (in part) on observations made with the Spitzer Space Telescope, which is operated by the Jet Propulsion Laboratory, California Institute of Technology under a contract with NASA. This research made use of Astropy, a community-developed core Python package for Astronomy \citep{Astropy2013}, NumPy and SciPy \citep{Oliphant2007}, Matplotlib \citep{Hunter2007}, {IPython} \citep{Perez2007} and NASA's Astrophysics Data System Bibliographic Services.

%%%%%%%%%%%%%%%%%%%% BIBLIOGRAPHY  %% %%%%%%%%%%%%

%\begin{thebibliography}{99}
\bibliographystyle{mn2e}                      % The reference style
\bibliography{fmr_z2}       % Multiple bib files.
%\include{../library.bib}
%\end{thebibliography}

\label{lastpage}

\bsp

\end{document}